\documentclass{aa}  
\usepackage{graphicx,lscape,deluxetable,longtable,natbib}
\usepackage{txfonts}
\def\ls{{_<\atop^{\sim}}}   
\def\gs{{_>\atop^{\sim}}}   
\def\cgs{ ${\rm erg~cm}^{-2}~{\rm s}^{-1}$ }    
\def\es{ ${\rm erg~\rm s}^{-1}$ }    
\def\cm{ ${\rm cm~\rm}^{-2}$ }    
\def\gtrsim{\mathrel{\hbox{\rlap{\hbox{\lower4pt\hbox{$\sim$}}}\hbox{$>$}}}}

\begin{document}
\title{The cosmological properties of AGN in the XMM-{\em Newton} Hard Bright Survey}


   \author{R. Della Ceca\inst{1}
        \and
	  A. Caccianiga\inst{1}
        \and
	 P. Severgnini\inst{1}
        \and
	 T. Maccacaro\inst{1}
        \and
	 H. Brunner\inst{2}
        \and
	 F. J. Carrera\inst{3}
        \and
	 F. Cocchia\inst{1,4}
        \and
	 S. Mateos\inst{5}
        \and
	 M.J. Page\inst{6}
        \and
	J.A. Tedds\inst{5}
	  }
	  
\offprints{R. Della Ceca}

   \institute{Istituto Nazionale di Astrofisica (INAF), Osservatorio Astronomico di Brera, Via Brera 21, 20121 Milano, Italy \\
          \email{roberto.dellaceca@brera.inaf.it}
         \and
          Max-Planck-Institut f\"ur extraterrestrische Physik, 
	  Giessenbachstrasse, 85741 Garching, Germany 
	 \and
          Instituto de F\'\i sica de Cantabria (CSIC-UC), Avenida de los
          Castros, 39005 Santander, Spain
         \and	
         INAF - Osservatorio Astronomico di Roma, via di Frascati 33, 00040 
	 Monte Porzio Catone, Italy 
         \and
         X-ray \& Observational Astronomy Group, Department of Physics and Astronomy, 
         Leicester University, Leicester LE1 7RH, UK
         \and
         Mullard Space Science Laboratory, University College London, 
         Holmbury St. Mary, Dorking, Surrey, RH5 6NT, UK
	 }
        
   \date{Received: December 26, 2007; accepted: April 28, 2008}

 
\abstract
{}
{We investigate here the  X-ray luminosity function (XLF) of absorbed ($N_H$ 
between $4\times 10^{21}$ and  $10^{24}$  cm$^{-2}$) and unabsorbed 
($N_H < 4\times 10^{21}$ cm$^{-2}$) AGN, the fraction of absorbed AGN as a
function of $L_X$ (and z), the intrinsic $N_H$ distribution of the AGN
population, and the XLF of Compton Thick ($N_H > 10^{24}$ cm$^{-2}$) AGN.}
{To carry out this investigation we have used the XMM-Newton Hard Bright
Serendipitous Sample (HBSS)  a complete sample  of bright X-ray sources   ($f_x
\gtrsim 7\times 10^{-14}$ erg cm$^{-2}$ s$^{-1}$) at high galactic latitude
($|b| > 20^{\circ}$) selected in the 4.5-7.5 keV energy band.  The HBSS sample is now
almost completely identified (97\% spectroscopic identifications)  and it can
be  safely used for a statistical investigation. The HBSS contains 62 AGN out of
which 40 are  unabsorbed (or marginally absorbed; $N_H  < 4\times 10^{21}$
cm$^{-2}$) and 22 are absorbed ($N_H$ between $4\times 10^{21}$ and 
$\sim 10^{24}$ cm$^{-2}$).}
{Absorbed and unabsorbed AGN are characterized by two different XLF with the
absorbed AGN population being described by a steeper XLF, if compared with the
unabsorbed ones, at all luminosities.  The intrinsic fraction F of absorbed AGN
(i.e. the fraction of sources with $N_H$ between $4\times 10^{21}$ and 
$10^{24}$ cm$^{-2}$ divided the sources with $N_H$ below $10^{24}$ cm$^{-2}$, 
corrected for the bias due to the photoelectric absorption)  with  $L_{2-10 keV}
\gtrsim 3\times 10^{42}$ \es  is $0.57\pm 0.11$;  we find that F  decreases with
the intrinsic luminosity, and probably, increases with the redshift. Our data
are consistent with a flat Log $N_H$ distribution for $N_H$  between $10^{20}$
and  $10^{24}$  cm$^{-2}$. Finally, by comparing the results obtained
here with those obtained using an optically selected sample of AGN we derive, in
an indirect way,  the XLF of Compton Thick AGN; the latter
is well described  by a XLF similar, in shape, to that of absorbed  AGN but
having a normalization about a factor 2 above. The density ratio between 
Compton Thick AGN ($N_H \gs 10^{24}$  cm$^{-2}$) and Compton Thin AGN ($N_H \ls
10^{24}$  cm$^{-2}$) decreases from  $1.08\pm 0.44$ at $\sim 10^{43}$ \es  to 
$0.57\pm 0.22$ at $\sim 10^{44}$ \es  to  $0.23\pm 0.15$ at $\sim 10^{45}$ \es.}
{The results presented here on the anti-correlation between F and $-L_x$ are
fully consistent with the hypothesis of a reduction of the covering factor of the
gas as a function of the luminosity and are clearly inconsistent with the
simplest unified scheme of AGN. These results strongly support the recently
proposed radiation-limited dusty torus model although alternative physical 
models 
are also consistent with the observations.}


\authorrunning {R. Della Ceca et al.} 
\titlerunning {The cosmological properties of AGN in the HBSS}
\maketitle
%

\section{Introduction}

For many years the study of the cosmological and statistical properties (e.g.
luminosity  function, cosmological evolution, mean spectral properties) of 
Active Galactic Nuclei (AGN)  was principally limited to the  optical (see e.g.
\citealt{boyle1988}; \citealt{croom2004})  or soft X-rays 
(see e.g. \citealt{maccacaro1991}, \citealt{dellaceca1992},  \citealt{miyaji2001},
\citealt{hasinger2005}) bands and essentially dealt with complete and unbiased
samples of optically unabsorbed (type 1) AGN.
Indeed  the selection and identification of sources hosting obscured accreting
nuclei is a difficult task: in the optical domain the active nuclei appear very
dim and their luminosity could be comparable to that of their host galaxies,
while in the  soft  X-ray band (up to few keV) their selection is difficult
since even hydrogen column densities  ($N_H$) of the order of
$10^{21}$-$10^{22}$ cm$^{-2}$ strongly reduce the flux from the nuclear source. 

However, despite their elusiveness, obscured AGN are fundamental for our 
understanding of the Super-Massive Black Holes (SMBHs) history as the large 
majority of the energy density generated by accretion of matter in the Universe
seem to take place in obscured AGN (\citealt{fabian1998}), as testified by the 
integrated  energy density contained in the cosmic X-ray background   (XRB;
\citealt{setti1989}; \citealt{madau1994}; \citealt{comastri1995}; 
\citealt{gilli2007}). Even more important,  the recent discovery of quiescent
SMBH in the nuclei of non-active nearby galaxies with prominent bulges
(\citealt{kormendy1995},  \citealt{magorrian1998}), along with the presence of
scaling relations between the central BH mass and galaxy  properties (e.g. bulge
luminosity/mass  and velocity dispersion, \citealt{ferrarese2000})  strongly
suggest that AGN are leading actors in the  formation and evolution of galaxies
and, in general, of cosmic structures in the Universe (see \citealt{begelman2004}
and references therein). Missing a large fraction of the AGN population could
thus bias our  understanding of the evolution of cosmic structure in the
Universe.

Hard X-rays (photon energies between a few keV and $\sim$ 10 keV) can directly
probe AGN activity, since they are almost uncontaminated by star formation
processes at the X-ray luminosities of interest ($L_X \gtrsim 10^{42}$\es), and
they are sensitive to absorbed  AGN up to an  intrinsic absorbing  column density
of $N_H$$\sim$$10^{23.5-24}$\cm, that is to say, they detect all but the most
absorbed sources.  Thus, hard X-ray surveys in the  2-10 keV range,  as pioneered
by  ASCA and BeppoSAX  (\citealt{cagnoni1998}; \citealt{ueda1998}; 
\citealt{dellaceca1999};  \citealt{fiore1999}; \citealt{ueda2003}),  and now
routinely performed with {\it Chandra} and {\it XMM--Newton}  \footnote{A
compilation of the main extragalactic surveys executed with  {\it Chandra} and
{\it XMM--Newton} is reported at the web address 
http://cxc.harvard.edu/xraysurveys/},  provide among the most  complete and
unbiased samples of AGN  presently possible (although they still miss
Compton Thick sources i.e. the sources with absorbing column densities
$\gs 10^{24}$\cm). In spite of this, a significant fraction of the AGN found in medium and deep
fields is too faint to provide good X--ray spectral information. Furthermore,
the extremely faint magnitudes of a large number of optical counterparts of
these X-ray faint sources make the spectroscopic identifications very difficult,
or even impossible, with the present day  ground--based optical telescopes.  

Here we use the bright ($f_X \gs$ 10$^{-13}$ \cgs), hard (4.5-7.5 keV selection
band),  almost completely identified (97\% spectroscopic ID) XMM-Newton Hard
Bright Sample,  to discuss a few issues which  are currently subject of intense
research activity, namely: a) the statistical properties (e.g. X-ray  luminosity
functions) of absorbed  and
unabsorbed AGN; b) the intrinsic ratio of absorbed and unabsorbed AGN as a
function of the X-ray luminosity; c) the intrinsic  absorption column
density distribution of the  AGN population for Hydrogen column density up to
$\sim 10^{24}$ cm$^{-2}$ 
and d) the X-ray luminosity function of Compton Thick AGN.
These are fundamental issues both to study and to follow
the accretion history in the Universe and to test the unification models of AGN.
We stress that the high identification rate of the HBSS sample is fundamental to
investigate these issues  since interesting and important  classes of X-ray
emitting sources (e.g. the type 2 QSOs) could be more difficult to identify and
therefore could be under-represented even in samples with an identification rate
of the order of 90\%.  Furthermore,  good optical and X-ray spectral data are
available for almost all the sources in the HBSS sample. In particular the
results of a complete X-ray  spectral analysis  will be efficiently used here 
to separate
absorbed AGN from the unabsorbed ones and  to compute k-corrections.

This paper is organized as follows: in \S \ref{par2} we give a short account of
the XMM-Newton Bright Serendipitous Survey and a description of the AGN sample
selected in the  4.5-7.5 keV range (the HBSS AGN sample hereafter).  In \S
\ref{par3} we study the cosmological evolution properties of the HBSS AGN sample
and derive the X-ray luminosity function (XLF) of the total AGN population as
well as the XLF of absorbed and unabsorbed AGN separately.  The intrinsic
fraction of absorbed  AGN as a function of the X-ray luminosities  is discussed
in  \S\ref{par4}, while in \S \ref{par5} we discuss the intrinsic N$_H$
distribution of the AGN population.   In  \S \ref{par6} we compare our results
with the  prediction of AGN unification models. In \S \ref{par7} we
compare  our results on the intrinsic fraction of absorbed AGN as a function of 
luminosity with those obtained using an optically selected sample of AGN;  this
comparison will be used to estimate the luminosity function of Compton Thick
AGN and to derive the ratio between Compton Thick and Compton Thin AGN as a 
function of $L_X$. 
Finally, summary and conclusions are reported in \S \ref{par8}.  
In Appendix A we discuss the method used to take into account the  photoelectric
absorption in the computation of the luminosity functions.  Throughout this
paper we consider the cosmological model with
($H_o$,$\Omega_M$,$\Omega_{\lambda}$)=(65,0.3,0.7);  results from other papers
have been rescaled to this cosmological framework.

\section{The XMM-Newton Bright Serendipitous Survey and the HBSS AGN Sample}
\label{par2}

The XMM-Newton Hard Bright Serendipitous Sample used here is part of a
bigger survey project known as XMM-Newton Bright Serendipitous Survey 
\footnote{The XMM-Newton Bright Serendipitous Survey is one of the
research programs conducted by  the XMM-Newton Survey Science Center (SSC,  see
http://xmmssc-www.star.le.ac.uk.) a consortium of 10 international institutions,
appointed by ESA to help the SOC in developing the software analysis system, to
pipeline  process all the XMM-Newton data, and  to exploit the XMM serendipitous
detections. The {\it Osservatorio Astronomico di Brera} is one of the Consortium
Institutes.} 
(XBS hereafter).
This latter consists of two flux-limited serendipitous samples of X-ray selected
sources at high galactic latitude ($|b| >20^{\circ}$):  the XMM BSS sample (389
sources) and the XMM HBSS sample (67 sources, with 56 sources in common with
the BSS sample) having an EPIC MOS2 count rate
limit, corrected for vignetting,  of $10^{-2}$ cts/s ($2\times 10^{-3}$ cts/s)
in the 0.5--4.5 keV (4.5--7.5 keV) energy band; the flux limit is  $\sim 7
\times 10^{-14}$ erg cm$^{-2}$ s$^{-1}$ in both energy selection bands. At the
time of this writing the spectroscopic identification rate is 87\% for the BSS
sample and 97\% for the HBSS sample.
The details on the XMM-Newton fields selection strategy and the source
selection  criteria of the XMM BSS and HBSS samples are discussed in 
\cite{dellaceca2004} while a description of the optical data and analysis,  of
the optical classification scheme and of the optical  properties of the
extragalactic sources identified  so far  is  presented in
\cite{caccianiga2008}. The  optical and X-ray properties of the galactic
population are discussed in \cite{lopezsantiango2007}.

Since the HBSS sample is now almost completely identified  it will  be used in
this paper for a statistical investigation. We note that among the ongoing
surveys performed with {\it Chandra} and XMM-Newton, the HBSS  is currently 
covering one of the  largest area ($\sim$ 25 deg$^2$) and,  unlike deep pencil
beam surveys, is unbiased by problems connected to the cosmic variance. The
current classification breakdown of the HBSS sample is as follows: 62  AGN, 1
cluster of galaxies (XBSJ 141830.5+251052) and 2 X-ray emitting stars (XBSJ
014100.6-675328 and  XBSJ 123600.7-395217). Two X-ray sources (XBSJ
080411.3+650906 and  XBSJ 110050.6-344331) are still unidentified at the time of
this writing. The redshift of the 62 AGN are reported in Table 1 along
with their basic X-ray spectral  properties that will be used in this paper
(e.g. photon index, intrinsic absorbing column density $N_H$, intrinsic 2-10 
keV  luminosity). A full account of the X-ray spectral properties for all the
sources in the XMM-{\it Newton} Bright Survey will be reported in a forthcoming
paper.

\subsection{Absorption properties}

\begin{figure}
   \centering
\includegraphics[width=9cm]{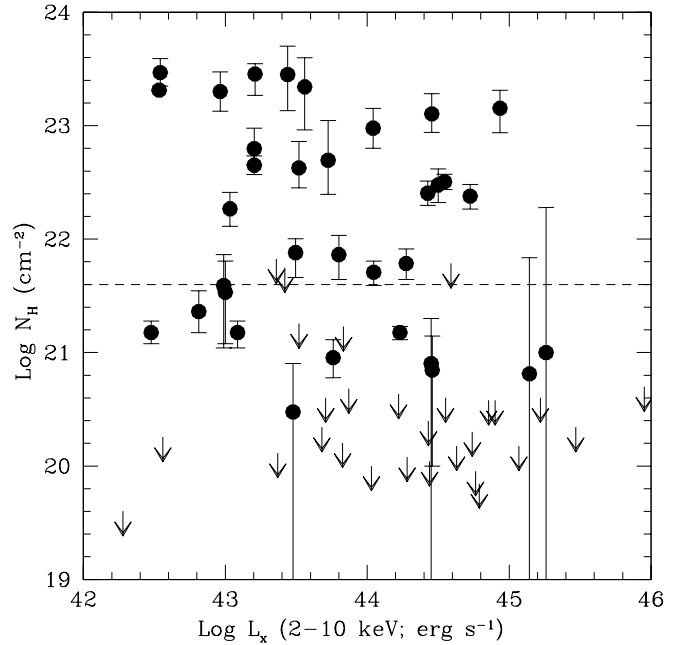}    
  \caption{Intrinsic 2-10 keV luminosity versus intrinsic absorption 
        column densities $N_H$ (both quantities derived from a 
	complete X-ray spectral analysis) for the AGN belonging to the 
	HBSS sample (filled circles: objects for which we have  
	measured the  $N_H$; downward arrows: objects for which we have only 
	an upper limit on $N_H$).}
         \label{lx-nh}
   \end{figure}

The intrinsic 2-10 keV luminosities versus intrinsic absorption  column
densities for the HBSS AGN sample are shown in Figure ~\ref{lx-nh}.  We
have directly measured the intrinsic absorbing column density for 34 AGN while
for the remaining 28 AGN we have only an upper limit.  However the measured 
upper limits are such that we can easily divide the sample into absorbed and
unabsorbed AGN using a value of $N_H  = 4\times 10^{21}$\cm. 
Assuming a Galactic A$_V$/$N_H$ ratio of $5.27\times 10^{-22}$ mag \cm 
(\citealt{bohlin1978}), $N_H = 4\times 10^{21}$ \cm corresponds to A$_V  \sim
2$; as shown and discussed in  \cite{caccianiga2008} (e.g. see their Figure 6) 
an  A$_V$ value of $\sim 2$ mag seems to be the best dividing value between 
optically type 1  (Sey1, Sey1.5, QSO, NLSy1) and
optically  type 2 (Sey1.8, Sey1.9, Sey2, QSO2) AGN in the XBS survey.
Therefore, the choice of $N_H = 4\times 10^{21}$ \cm as dividing line between 
X-ray absorbed and unabsorbed AGN has a direct physical link to the optical
classification.
Three objects (XBS J041108.1-711341, XBS J205635.2-044717 and XBS 
J220601.5-015346)   have an upper limit between  $N_H  = 4\times 10^{21}$ \cm
and $N_H \sim 7\times 10^{21}$ \cm  (see Figure ~\ref{lx-nh}), so the current X-ray
data does not allow a firm classification as absorbed or unabsorbed AGN. On the
other hand these three objects are classified as broad line (Type 1) AGN in the
optical domain (\citealt{caccianiga2008}); based on their optical properties
these three objects will be considered here as unabsorbed  AGN.  With these
choices the sample of unabsorbed (or marginally absorbed) AGN is composed of 40
objects while the sample of absorbed AGN is composed of 22 objects.
As shown in Section 4 the specific choice of the dividing line between
absorbed and unabsorbed AGN does not affect the main results reported 
in this paper.

For the sake of clarity we will call in this paper as  {\it unabsorbed AGN} 
those AGN with an intrinsic absorbing column density $N_H$ below  $4\times
10^{21}$  cm$^{-2}$, as {\it absorbed AGN} those objects with  $N_H$  between
$4\times 10^{21}$ and  $\simeq 10^{24}$  cm$^{-2}$ and as {\it Compton Thick AGN}
those sources with $N_H$ above $10^{24}$  cm$^{-2}$.

%
%

At the flux limit of the HBSS survey the measured surface densities of absorbed
(unabsorbed) AGN  is $0.87^{+0.23}_{-0.18}$  ($1.59^{+0.25}_{-0.25}$) deg$^{-2}$
and the observed  fraction of absorbed AGN is  $37\pm 7 \%$.  
If we split the sample according to an intrinsic luminosity of  $10^{44}$ \es we
have that the fraction of absorbed AGN is  $29^{+10}_{-9}\%$ in the high
luminosity regime and $42^{+11}_{-10}\%$  in the low luminosity regime, thus
confirming  first results based on a complete sub-sample of 28 HBSS objects
fully identified reported in \cite{caccianiga2004}.  However this crude
comparison  between low and high luminosity AGN is biased since it  does not
take into  account selection effects  due to the absorption, i.e. we can observe
unobscured sources within a larger volume compared to the obscured ones  since
the intrinsic luminosity of the latter sources is depressed by 
photoelectric absorption.  We will derive and discuss the unbiased intrinsic
fraction of absorbed AGN as a function of the intrinsic X-ray luminosity in
section 4.

It is worth noting that the selection strategy of the HBSS survey is extremely
efficient in finding type 2 QSO, i.e. absorbed AGN with an intrinsic luminosity 
above $10^{44}$\es. We have 9 of these sources in the HBSS sample (7 with  $N_H >
10^{22}$ cm$^{-2}$); they represent $15^{+6}_{-5}\%$  of the total AGN population
and  $41^{+13}_{-12}\%$  of the  absorbed ones.  At the flux limit of the HBSS survey
the measured surface density of type 2 QSOs is  $0.36^{+0.13}_{-0.12}$ deg$^{-2}$;
this latter surface density of  Type 2 QSO combined with the surface density of
Type 2 QSOs reported in  \cite{perola2004} ($\sim 48$ deg$^{-2}$ at a 2-10 keV flux
limit of  $\sim  10^{-14}$\cgs) implies a slope in the integral LogN-LogS of Type 2
QSOs of $\sim -2.1$ between $\sim  10^{-13}$  and $\sim 10^{-14}$\cgs.

Finally we have not found Compton Thick AGN, which are a fundamental population
to produce the shape of the XRB around its emissivity peak  ($\sim$ 30 keV,
e.g.  \citealt{gilli2007}). As discussed in  Appendix A even the 4.5-7.5
keV HBSS survey is quite inefficient in selecting this kind of AGN. Given their
large flux depression due to the absorption, X-ray surveys above 10 keV are
needed to select  this kind of sources  (see \citealt{dellaceca2007} for a
recent review of this topic). We discuss Compton Thick AGN in Section 7
where we derive, in an indirect way, their X-ray luminosity function.

\subsection{Redshift and Luminosity distribution}
 
\begin{figure}
   \centering
\includegraphics[width=7cm]{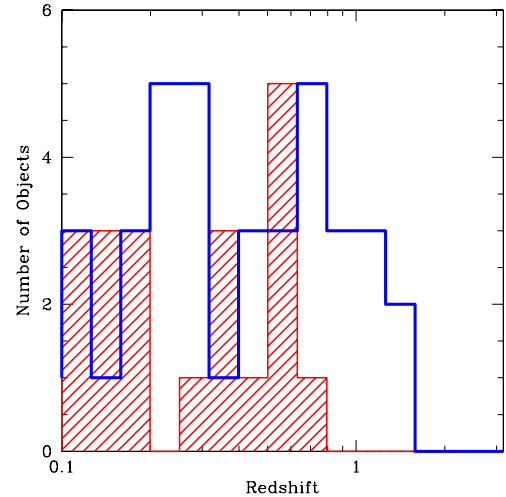}    
  \caption{Redshift distribution of the AGN belonging to 
           the HBSS sample (shaded histogram: absorbed AGN; 
	   empty histogram: unabsorbed AGN) 	   
          }
         \label{histo-z}
   \end{figure}

\begin{figure}
   \centering
\includegraphics[width=7cm]{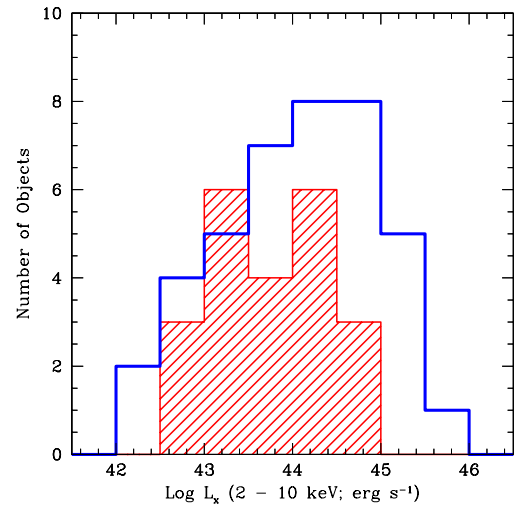}    
  \caption{Intrinsic 2-10 keV luminosity distribution of the 
           AGN belonging to 
           the HBSS sample (shaded histogram: absorbed AGN; 
	   empty histogram: unabsorbed AGN)	   	   
              }
         \label{histo-lx}
   \end{figure}

The redshift and intrinsic 2-10 keV luminosity distribution for the AGN
belonging to the HBSS sample are reported in Figure ~\ref{histo-z} and
Figure ~\ref{histo-lx}, respectively.  The shaded histogram indicates the
distribution for absorbed  AGN while the empty  histogram indicates the
distribution for the unabsorbed ones. Both classes of AGN show a rather flat
{\it z} distribution with the unabsorbed AGN population sampled up to $z\sim
1.5$  while the absorbed AGN population is sampled up to $z\sim 0.8$. Intrinsic
2-10 keV luminosities are distributed over 4 order of magnitude  for the
unabsorbed AGN population (from $10^{42}$ to $10^{46}$ erg s$^{-1}$  with a
median logarithmic luminosity of $\sim 44.2$) and over about  2.5 order of 
magnitudes for the absorbed AGN population (from $\sim 7\times 10^{42}$ to 
$10^{45}$ erg s$^{-1}$  with a median logarithmic luminosity of 43.7).

%

%

\section{Cosmological properties}
\label{par3}

\subsection{Evolution}

For the sample of unabsorbed AGN (40 objects) we find that, according to the
$V_e/V_a$ test (\citealt{schmidt1968}; \citealt{avni1980}),  the hypothesis of a
uniform distribution of the objects in the Universe  is rejected at a confidence
level of $\sim 98\%$  ($<V_e/V_a> = 0.616 \pm 0.046$).  Assuming a pure
luminosity evolution model with an evolutionary form  $\propto (1+z)^C$ we find
a best fit parameter of C$\simeq$2.7 with an associated 68\% confidence interval
of 1.9-3.0.  
This value of cosmological evolution is consistent, within the errors, both with
the results obtained in the soft (E$\ls 3$ keV) energy band using the {\it
Einstein} Extended Medium Sensitivity Survey ($C=2.56\pm 0.17$;  
\citealt{maccacaro1991}, \citealt{dellaceca1992}) and the EMSS+Rosat AGN samples
($C=2.6\pm 0.1$; \citealt{page1997}) and with the results in the 2-10 keV energy
range reported in  \cite{ueda2003} ($C=2.70^{+0.17}_{-0.25}$) and
\cite{lafranca2005} ($C=3.22^{+0.13}_{-0.26}$).

However it is now well established that a  Luminosity Dependent Density
Evolution (LDDE) model provides a better description of the
evolutionary  properties of AGN, both in the X-ray  energy
range (\citealt{hasinger2005};  \citealt{ueda2003}; \citealt{lafranca2005};
\citealt{silverman2007}) and in the optical domain  (\citealt{bongiorno2007}).

To test this evolutionary behavior, we assume here  an LDDE model with the
parametrization  as introduced by \cite{ueda2003}, 

$${d \Phi(L_x,z) \over d Log L_x}  = {d \Phi(L_x,0) \over d Log L_x} \times
e(z,L_x)$$

$$ e(z,L_x) =
\left\{
\begin{array}{lr}
(1+z)^{p1}, &   (z<z_{\rm c}) \\
 e(z_{\rm c},L_x)[(1+z)/(1+z_{\rm c})]^{p2}, &    (z \geq z_{\rm c})
\end{array}
\right.
$$

$$ z_{\rm c}(L_{\rm x}) =
\left\{
\begin{array}{lr}
z_{\rm c}^*,  & (L_{\rm X} \geq L_a) \\
z_{\rm c}^*(L_{\rm X}/L_a)^\alpha  & (L_{\rm X}<L_a).
\end{array}
\right.
$$

where z$_{\rm c}$ corresponds to the redshift where the direction of the 
evolution changes sign.  It is worth noting that z$_{\rm c}$ is a function of the
intrinsic luminosity  of the object;  if we assume the best fit parameters of 
p2=-1.15; z$_{\rm c}^*=2.49$; $\alpha$=0.20; Log L$_a$=45.80 (adapted to 
 H$_0$=65) as reported in 
\cite{lafranca2005},  then z$_{\rm c}$ is $\sim 0.7,1.1,1.7$  for AGN  with
L$_{\rm X}$ $\sim 10^{43}, 10^{44}, 10^{45}$ \es, respectively.  Given the
coverage in the luminosity-redshift plane of the HBSS unabsorbed AGN 
sample, for each luminosity
the objects are below  z$_{\rm c}$, implying that we are unable to derive p2,
z$_{\rm c}^*$, $\alpha$ and  LogL$_a$.  For this reasons we have fixed them from
\cite{lafranca2005} and we have used the $V_e/V_a$ test to constrain p1. 

We obtain a best fit p1=6.5 with  an associated 68\% confidence interval of 3.5
- 10.0.  The distribution of the derived $V_e/V_a$ values is consistent with
being  uniformly distributed between 0 and 1  according to a KS test (KS
probability $\sim 95\%$). We have also checked that, given the coverage of the
luminosity-redshift  plane  of the HBSS AGN sample, the best fit  p1 is
virtually  insensitive to the other parameters of the model;  i.e., p1 does not
change by varying all the  other parameters within their $1\sigma$  range as
derived from  \cite{lafranca2005}.

The derived best fit value for p1 is consistent, within the errors,  with
that reported in \cite{lafranca2005} (p1=$4.62\pm 0.26$) and is in  very good
agreement with those recently obtained, in the optical domain, by 
\cite{bongiorno2007} using a sample of 130 broad line AGN with redshift  up to
z=5 from the  VIMOS-VLT Deep Survey (p1=6.54) and from \cite{hopkins2007}  using
a large data set of AGN selected in the Mid-IR,  optical, soft X-ray and hard
X-ray (p1=$5.95\pm 0.23$).

Because of their  number statistics (22 objects in total) and their distribution in the 
$L_X-z$ plane  the cosmological evolution is unconstrained for the absorbed AGN
sample (note that the absorbed AGN are sampled only  up to z$\sim 0.8$).
Therefore in the following, and in line with the Unification Scheme of AGN, we
will make the assumption that this class of sources evolve with cosmic time 
(and within the reshift range sampled at the HBSS flux limit) in a similar way
as the unabsorbed ones.

\subsection{X-ray Luminosity Functions of Absorbed and Unabsorbed AGN}

In this section we will derive the {\it intrinsic} (i.e. corrected for the
selection bias due to the intrinsic absorption) local X-ray luminosity function of
absorbed  and unabsorbed AGN.  For comparison with other survey projects the
XLF will be derived in the 2-10 keV energy range using the $1/V_a$ method
(\citealt{avni1980}) as follows.

For each AGN  we have computed the maximum redshift ($z_{max}$) at  which the
source (characterized by the observed  count rate in the 4.5-7.5 keV energy
band, by the X-ray spectral shape  and by the redshift) can still  be detected
above the MOS2 count rate limit of the  HBSS sample ($=2\times 10^{-3}$ cts/s in
the  4.5-7.5 keV band).  The X-ray spectral shape of each AGN, as derived from
the X-ray spectral analysis (reported in Table 1), has been used to take
properly into account k-corrections.  

This $z_{max}$  combined with the solid angle covered by the HBSS survey
(25.17 deg$^{-2}$) and the best fit LDDE model described  in the previous
section allow us to compute, for each  object $i$, the {\it density-weighted
volume}  $V_{a}^{\prime}$ (see \citealt{avni1980})
$$ V_{a_i}^{\prime} = 25.17\times \int_{0}^{z_{max}} e(z,L_x) \times dV/dz
\times dz $$
The XLFs computed using these volumes are the so called de-evolved XLF at 
z=0.

In order to obtain the differential 2-10 keV X-ray luminosity function  we bin
the individual contributions in bins of equal  logarithmic width $\Delta Log L_X$
according to the intrinsic 2-10 keV  luminosity as derived from the X-ray
spectral analysis of each source. For each luminosity bin we have  $$ {d \Phi
\over d Log L_x}  = \sum_{i=1}^{n} [{1\over V_{a{_i}}^{\prime} \Delta Log
L_x}]$$ where n is the number of objects in that bin.  Poissonian errors bars
have been computed using the \cite{gehrels1986} prescriptions.

In summary, in the XLF computation we have used:
 
a) the observed count rate in the selection 4.5-7.5 keV band, the count rate limit
of the HBSS survey and the best fit LDDE model in estimating the {\it available}
co-moving density weighted volume $V_{a_i}^{\prime}$ of each source. The spectral
shape of each individual source, as derived from the X-ray spectral analysis, has
been used to properly  take into account k-corrections; 

b) the intrinsic (2-10 keV) luminosities (as derived from the X-ray spectral 
analysis of each source) to divide the sources into luminosity bins.

In Appendix A  we discuss a few simulations we have carried out to prove the 
validity of this approach.

\addtocounter{table}{+1}
\begin{table*}
\begin{center}
\caption{Best Fit Parameters of the 2-10 keV XLF}
\begin{tabular}{rrrrrrr} 
\hline \hline
Sample           & Objects  & Log A                         & Log $L_{\star}$   &  $\gamma_1$           &  $\gamma_2$    & KS-prob \\
(1)               & (2)      & (3)                           & (4)              &  (5)                 &   (6)         &  (7)    \\
\hline 
Unabsorbed AGN   & 40       & $-6.21^{+0.12}_{-0.12}$      & 44.0 (fixed)      & $1.08^{+0.19}_{-0.19}$ & $2.38^{+0.17}_{-0.18}$ & 0.44; 0.24; -\\
Absorbed AGN     & 22       & $-6.62^{+0.17}_{-0.16}$      & 44.0 (fixed)      & $1.55^{+0.31}_{-0.24}$ & $2.61^{+0.45}_{-0.52}$ & 0.10; 0.49; 0.34 \\
\hline \hline
\end{tabular}
\end{center}
Columns are as follows: 
(1) Sample;
(2) number of objects in the sample;
(3) XLF normalization, A, is in units of  $h^3_{65}$ $Mpc^{-3}$; 
(4) XLF break luminosity $L_{\star}$ in units of $h^{-2}_{65}$
    \es. Since $L_{\star}$ is not well constrained we have fixed it to
    $10^{44}$  \es  close to that found for the total AGN population  in
    \cite{sazonov2004}  ($L_{\star}=10^{43.6}$ \es) and in \cite{shinozaki2006}
    ($L_{\star}=10^{44.02}$ \es);
(5) low-luminosity XLF slope;
(6) high-luminosity XLF slope;
(7) the three values are the probabilities (1D-KS test) that the observed z,
    L$_x$ and, in the case of absorbed AGN, $N_H$ distributions
    are consistent with being derived from the the best fit XLF plus 
    cosmological evolution plus intrinsic $N_H$ distribution derived here.
    See section 3.2 for details.   
\end{table*}

   \begin{figure}
   \centering
\includegraphics[width=9cm]{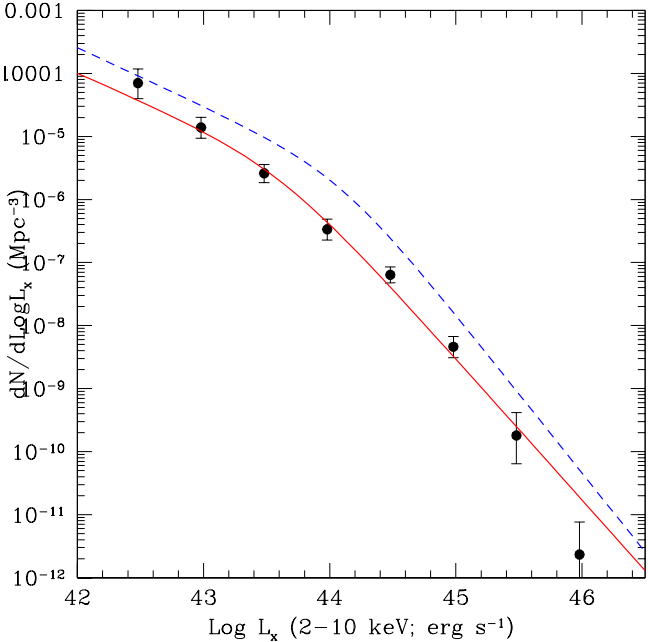}    
  \caption{De-evolved (z=0) X-ray luminosity function (2-10 keV)  obtained
	   using  the total HBSS AGN sample (filled circles). 
	   We have also reported for comparison  the local  XLF for the
	   total AGN population  as derived by  \cite{shinozaki2006} (HEAO1
	   AGN sample; dashed line) and by  \cite{sazonov2004}
	   (RXTE AGN sample; solid line). See section 3.2 for details.
	   }
         \label{XLF_1}
   \end{figure}
%
%

In Figure \ref{XLF_1}  we show (filled dots) a binned representation of the
de-evolved (2-10 keV) XLF  obtained using  the total AGN HBSS sample (62
objects). The derived  XLF (at z=0) is compared with the 2-10 keV XLF obtained 
using {\it local} AGN samples, the latter being independent from cosmological
evolution. 
In particular in Figure \ref{XLF_1} we compare
our XLF with that obtained from \cite{shinozaki2006} using the 2-10 keV selected
AGN in the HEAO1 survey (49 objects in total, dashed line) and by 
\cite{sazonov2004} using the 3-20 keV selected AGN sample (95 AGN in total;
solid line) in the Rossi X-ray  Timing Explorer (RXTE) slew survey. This last
XLF has been converted from the 3-20 keV to the 2-10 keV energy range using a
power-law model with photon index equal to 1.9 (${L_{2-10 keV} \over L_{3-20
keV}} = 0.8$). As can be seen in Figure \ref{XLF_1} the best fit AGN XLF (at
z=0)  derived here is in excellent agreement with that derived using the RXTE 
slew data while its normalization is a factor between 2.5 to 5 times below that
derived using the  HEAO1 AGN sample (\citealt{shinozaki2006}). This discrepancy
in the normalization of the local (2-10 keV) XLFs was already  discussed by
\cite{sazonov2004} and by \cite{shinozaki2006} and it is currently  an open
issue.   It is worth noting that a similar shift between our computation of the
XLF and the \cite{shinozaki2006} XLF is also present when we consider only  the
unabsorbed AGN population, suggesting that the possible different distribution 
of the absorption properties  of the two  samples cannot explain the different
observed normalizations of the XLFs.  Despite the different normalizations the
shapes of the different XLFs seem to  be in satisfactory agreement.

\begin{figure}
   \centering
\includegraphics[width=9cm]{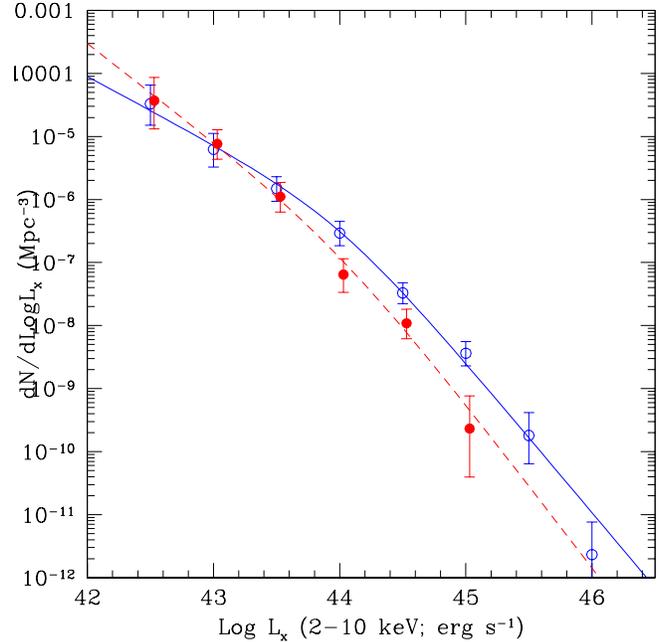}    
  \caption{Comparison between the de-evolved (z=0) X-ray luminosity function 
	   (2-10 keV)  for the unabsorbed 
	   ($N_H$ $<4\times 10^{21}$ cm$^{-2}$; open circles) 
	   and 
	   absorbed ($4\times 10^{21}< N_H \ls 10^{24}$ cm$^{-2}$; 
	   filled circles) AGN in the HBSS sample. 
	   The solid and dashed lines represent
	   the best fit two power-law function of the unabsorbed and absorbed 
	   AGN XLF, respectively (best fit parameters as reported in Table 2).
           Please note that for comparison reasons the XLF of absorbed AGN 
	   has been computed with a small shift ($\Delta$ Log L$_x$=0.03) with 
	   respect to that of unabsorbed AGN.
	   See section 3.2 for details.}
         \label{XLF_2}
   \end{figure}

More important is the comparison between the XLF of the absorbed and  unabsorbed
AGN population in the HBSS sample reported in Figure ~\ref{XLF_2}  (unabsorbed
population: open circles; absorbed population: filled circles).  It is clear
that the two populations are characterized by two different XLF with the
absorbed AGN population being described by a steeper XLF, if compared with the
unabsorbed ones, at all luminosities.  The binned representation of the XLFs
reported in  Figure \ref{XLF_2} have been fitted by a smoothly connected two
power-laws function of the form  $${d \Phi(L_x,z=0) \over d Log L_x}  = {A [
({L_x \over L_{\star}})^{\gamma_1} + ({L_x \over L_{\star}})^{\gamma_2}]^{-1}}$$
taking into account the error bars of each data point and    by minimizing
$\chi^2$ using the routines in the QDP  \footnote{See in
http://heasarc.gsfc.nasa.gov/docs/software}  software package. Best fit XLF
parameters and 1 $\sigma$ errors are reported in Table 2.

%
%
    
For the unabsorbed AGN population the value of $\gamma_1$ and  $\gamma_2$  are
consistent, within the errors, with those reported in  \cite{sazonov2004} and in
\cite{shinozaki2006}.   For the absorbed AGN population the $\gamma_1$
($\gamma_2$) derived here  is slightly  steeper (flatter) than that reported in
\cite{shinozaki2006}  ($\gamma_1=1.12^{+0.17}_{-0.19}$;
$\gamma_2=3.34^{+0.90}_{-0.65}$).  \cite{sazonov2004} does not quote the XLF
parameters for the  absorbed AGN population.

As discussed in section 2 two HBSS sources (XBSJ080411.3+650906 and 
XBSJ110050.6-344331) are still unidentified at the time of this writing. Can their
inclusion in the AGN sample substantially change the results discussed above? 
In particular their inclusion could be important if these sources were  high
luminosity absorbed AGN where we measure a deficit of absorbed sources. 
To answer to this question we have examined in detail their X-ray spectral
properties as well as the optical (photometric) properties of the most likely
optical counterpart. XBSJ110050.6-344331 is well described (fixing z=0) by a
power-law model having  $\Gamma = 1.80\pm 0.16$ and $N_H=1\pm0.5 \times 10^{21}$
cm$^{-2}$.  Its 2-10 keV X-ray flux ($\sim 3 \times 10^{-13}$ \cgs), optical
magnitude (m$_R$ = 18.0),  X-ray to optical flux ratio ($X/O \sim 1$) and X-ray
spectral properties strongly suggest a type 1 (unabsorbed) object. On the
contrary  XBSJ080411.3+650906 is most likely an absorbed AGN. Its X-ray spectra
is described (at z=0) by a power-law model having $\Gamma = 1.7\pm 0.4$ and
$N_H=8.4\pm 4.3 \times 10^{21}$ cm$^{-2}$, so with  an intrinsic $N_H$ (at the
source z) well above $4\times 10^{21}$ cm$^{-2}$.  The optical magnitude of the
counterpart (m$_R$ = 21.10) combined with its 2-10 keV X-ray flux ($\sim 2.1
\times 10^{-13}$ \cgs) implies an X/O flux ratio of $\sim 17.5$, strongly
supporting the absorbed AGN hypothesis  (e.g. \citealt{severgnini2006}). If we
use the relationship between X/O flux ratio and  intrinsic luminosity reported in
\cite{fiore2003} the most likely redshift of this object is $\sim 0.55$ 
(consistent with a poor quality optical  spectrum of the optical counterpart)
and its intrinsic luminosity is  $L_x\simeq 1.9\times 10^{44}$ \es. The addition
of this object to the HBSS absorbed AGN sample produce an increase of $\sim 8\%$
of the space density of the  absorbed AGN population at $L_x\simeq 10^{44}$
\es, a very marginal difference that is well within the error bar reported  in
Figure \ref{XLF_2}. 
It is also worth noting that there are four objects (see Figure ~\ref{lx-nh}) which
have a best fit $N_H$ value below  $4\times 10^{21}$ cm$^{-2}$ but with error
bars  crossing this $N_H$ dividing line. The two objects (having large error
bars) with a luminosity around $2\times 10^{45}$ \es are optically classified as
broad line QSO; so their optical properties are fully consistent with the
unabsorbed classification. The two objects with  a luminosity around $1\times
10^{43}$ \es are instead optical elusive AGN (\citealt{caccianiga2007a}) and
their unabsorbed classification was based on their best fit $N_H$ value. On the
other end their  inclusion in the absorbed AGN sub-sample will increase the
difference between  absorbed and unabsorbed AGN around $L_x \simeq 10^{43}$ \es
strengthening our  conclusions of different XLF of the two populations.  To
conclude, the results presented and discussed in this section are stable  also
taking into account all the uncertainty about the two unidentified sources and
about our present classification (absorbed or unabsorbed AGN) break down.
Finally, as consistency test, we have compared the observed z, L$_x$ and  $N_H$
distributions with those predicted by the best  fit cosmological evolution model
derived here (XLF + cosmological evolution) folded with the  $N_H$  distribution
discussed in section 5. The 1D-KS probabilities reported in Table 2 confirms the
goodness of our modeling of the AGN cosmological properties.

\section{The intrinsic fraction of absorbed AGN}
\label{par4}

One of the most important open issues regarding  absorbed AGN is to understand their
relevance amongst the AGN population  and if (and how)  the fraction (F)  of absorbed
AGN  changes as a function of $L_X$ and z.  We define F as the intrinsic ratio
between the AGN with  $N_H$ in the range  $4\times 10^{21}$ -- $10^{24}$  cm$^{-2}$
and all the AGN with $N_H$ below $10^{24}$  cm$^{-2}$.  This is a long standing
issue coming back to almost 26 years ago when  \cite{lawrence1982} reported an
anti-correlation between the intrinsic  luminosity of the AGN and the presence of
absorption. This issue has been recently re-discussed in many papers  with
contradicting results: several studies suggest  that F  decreases with L$_x$ and
increase with z (e.g. \citealt{lafranca2005}; \citealt{treister2006}),   while other
studies suggest that F is independent of L$_x$ and z (e.g. \citealt{dwelly2006}) or 
that F is dependent from L$_x$  but not from z (e.g. \citealt{ueda2003};
\citealt{akylas2006}, \citealt{gilli2007}). Having computed the intrinsic de-evolved
XLFs of absorbed and unabsorbed  AGN  we can now use them to derive F.  We stress
that  using the derived XLFs we should be free from the selection bias due to the
absorption  since the appropriate  corrections have been already  taken into account
in the computation of the XLF itself. 
We also recall that we are not considering here Compton Thick AGN, so the
derived F ratio is indeed a lower limit to the {\it true} fraction of absorbed AGN.
Compton Thick AGN will be discussed in section 7. 

Using the integral representation of the derived XLFs  we have first computed F
for AGN having an intrinsic luminosity above $\sim 3\times 10^{42}$ \es (i.e.
around the luminosity of the faintest obscured AGN in the HBSS sample).  The 
fraction ($F=0.57\pm 0.11$) is in excellent agreement  with that found
using hard ($E>10$ keV) selected (local) samples of AGN from INTEGRAL/Swift
surveys  at a flux limit of  $\sim 10^{-11}$ \cgs.   This comparison is reported in
Table 3, where we have listed the value of  F in the HBSS sample, in the
SWIFT/BAT sample (\citealt{markwardt2005}, \citealt{ajello2008}) and in the
INTEGRAL samples  (\citealt{beckmann2006}; \citealt{bassani2006}). Please note
that the local INTEGRAL/Swift samples are almost free from selection bias
related to the absorption at least for $N_H$  up to $10^{24}$ cm$^{-2}$. The
good agreement implies that, after selection effect due to the absorption have
been taken  into account, the HBSS is sampling at a $f_X \sim 10^{-13}$ \cgs, 
the same population  of absorbed AGN as the present surveys
performed at $E>10$ keV. It is also worth noting that after selection effect
have been taken into account the fraction of obscured AGN increases  from an
observed value of $0.37\pm 0.07$ (see section 2.1) to an intrinsic ones of
$0.57\pm 0.11$.

\begin{table}
\begin{center}
\caption{Fraction of absorbed AGN with $L_x \gs 3\times10^{42}$ \es 
in different samples$^1$}
\begin{tabular}{rrr} 
\hline \hline
Sample                          & F                   & NOTE              \\
\hline 
HBSS                      & 0.57$\pm$ 0.11       &   XMM              \\
\cite{markwardt2005}        & 0.59$\pm$ 0.09       &   SWIFT/BAT        \\
\cite{beckmann2006}          & 0.54$\pm$ 0.10       &   INTEGRAL        \\
\cite{bassani2006}          & 0.56$\pm$ 0.09       &   INTEGRAL        \\
\cite{ajello2008}          & 0.57$\pm$ 0.13       &   SWIFT/BAT        \\
\hline 
\hline 
\end{tabular} 
\end{center} 

Notes:  
$^1$ In deriving, from the original papers, the above reported quantities we
have considered, as for the HBSS AGN sample, only the Compton  Thin AGN ($N_H <
10^{24}$ cm$^{-2}$) and a dividing $N_H$  line  between absorbed and
unabsorbed AGN fixed at  $4\times 10^{21}$ cm$^{-2}$. 
We have not reported the INTEGRAL results from \cite{sazonov2007}  since these
authors do not report the absorbing column densities of their objects but 
simply split the objects into  absorbed and unabsorbed according to an  $N_H$
value of  $10^{22}$ cm$^{-2}$; for comparison  the fraction of absorbed AGN 
found from their sample using this $N_H$ value is $0.42\pm 0.09$.
\end{table}
      
\begin{figure*}
\begin{center}
\begin{tabular}{lll}
\includegraphics[width=0.33\textwidth]{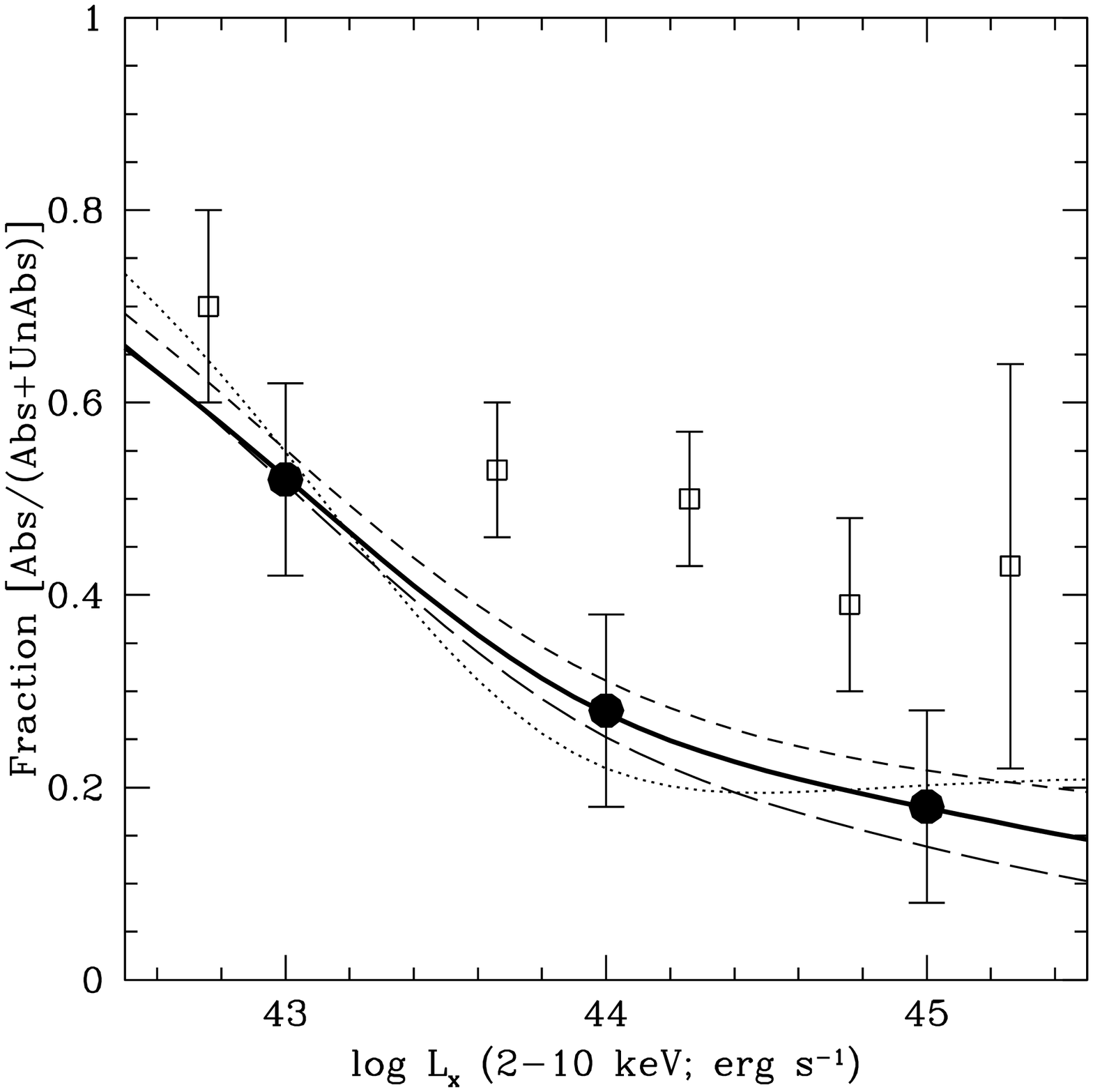}
&\includegraphics[width=0.33\textwidth]{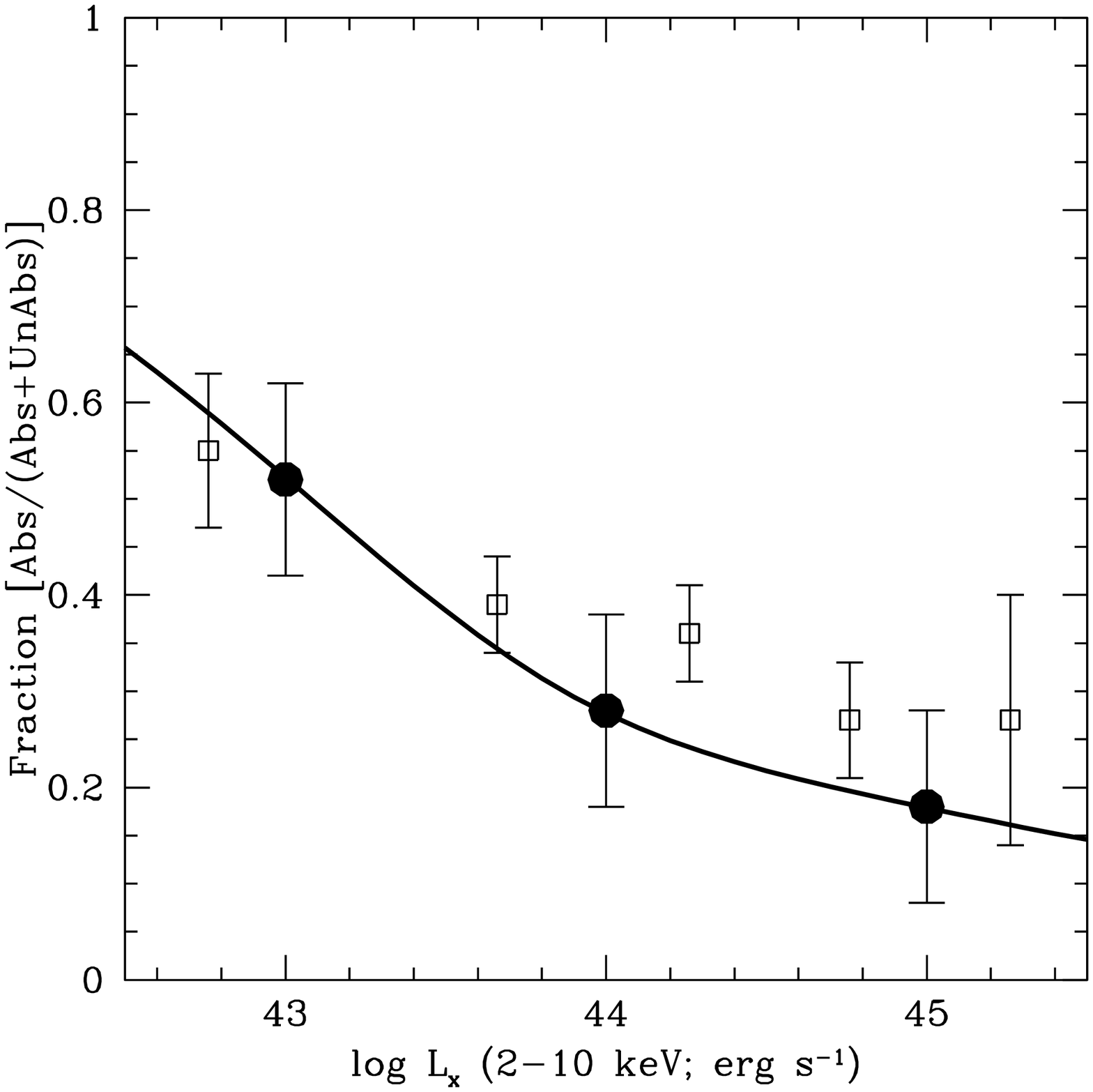}
&\includegraphics[width=0.33\textwidth]{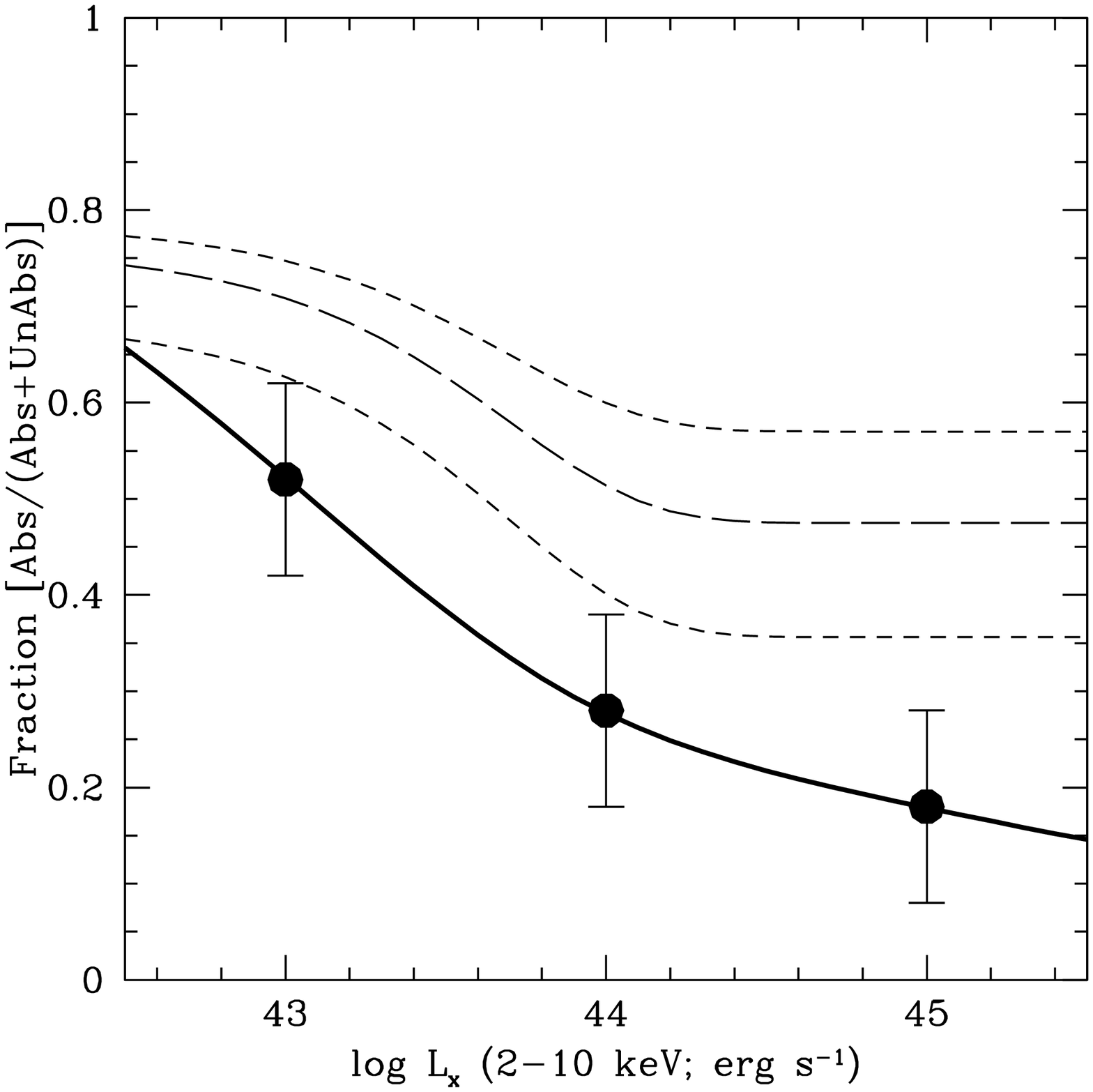}
\end{tabular}
\caption{
{\it Left panel}: the thick solid line represents the differential  fraction F of
absorbed AGN  as a  function of the intrinsic luminosity obtained using the best fit 
de-evolved XLF of  absorbed ($N_H$  between $4\times 10^{21}$ and  $10^{24}$ 
cm$^{-2}$)  and unabsorbed ($N_H$  below $4\times 10^{21}$ cm$^{-2}$) AGN in the HBSS
survey. The 68\%  confidence errors on this ratio are also reported at   $L_x =
10^{43}, 10^{44}, 10^{45}$ \es.  We have also reported the differential fraction of
absorbed AGN obtained assuming no cosmological evolution (short dashed line), a
cosmological  evolution of the  absorbed/unabsorbed ratio similar to that recently
proposed by  \cite{treister2006} (long dashed line) and that obtained using $N_H =
10^{22}$ cm$^{-2}$ as dividing value between absorbed and unabsorbed AGN (dotted
line). The open squares represent the fraction of absorbed AGN  as derived from
\cite{akylas2006}.  {\it Middle panel}: solid line as in the left panel.  The
\cite{akylas2006} points have been rescaled to z=0 using the mean redshift in each 
luminosity bin and an evolution of the  absorbed/unabsorbed ratio similar to that
recently proposed by \cite{treister2006}.   {\it Right panel}: solid line  as in the
left panel. The long dashed line  is the  prediction from \cite{gilli2007} on the
synthesis  model of the CXB. The two dashed lines enclose the  uncertainty on the F
ratio  based on the CXB modeling.  We stress that all the quantities reported in this
figure refer to AGN with  absorbing column density  $N_H$ below  $10^{24}$ cm$^{-2}$.
See  section 4 for details.}
\label{ratio}
\end{center}
\end{figure*}

The differential  fraction F of absorbed AGN as a function of the intrinsic
luminosity is reported in  Figure \ref{ratio} left panel (solid line);  we have
also reported the 1$\sigma$ error bar on F (vertical solid line at $L_x =
10^{43}, 10^{44}, 10^{45}$ \es),  computed taking into account the current 
uncertainties on the XLFs.

Given the coverage in the luminosity-redshift plane of the HBSS AGN sample the
results reported in Figure \ref{ratio} are quite stable,  i.e. they do not depend
strongly on the value of the cosmological evolution parameter. To prove this in
Figure \ref{ratio} (left panel) we have also reported F as a function of $L_X$  if we
assume no cosmological evolution for absorbed and unabsorbed AGN (short dashed line)
or if we assume a cosmological evolution  of the absorbed population faster than that
the unabsorbed ones according to the \cite{treister2006} model ($F \propto
(1+z)^{0.4}$; long dashed line). Finally we have also reported (dotted line) the
fraction of absorbed AGN if we re-perform all the  analysis but this time splitting
the objects as absorbed or unabsorbed using $N_H = 10^{22}$ cm$^{-2}$ as dividing
value (44 unabsorbed AGN, 18 absorbed AGN). 

In the same left panel of Figure \ref{ratio} we have also reported the fraction of
obscured AGN as  recently computed by \cite{akylas2006} using a sample of AGN
selected in the hard 2-8 keV band down to a flux limit of  $6\times 10^{-16}$
\cgs (open squares),  about a factor one hundred below our flux limit. The data
points computed by \cite{akylas2006} have been corrected for the different
volumes sampled from absorbed and unabsorbed AGN (due to the absorption) in a
similar way as done here. Although the HBSS sample and the \cite{akylas2006}
show a  similar trend of decreasing F as a function of the intrinsic luminosity,
the fractions computed at z=0 using the HBSS sample are systematically  below
($\sim$ a factor 2  for $L_x > 10^{43}$ \es) than those reported in
\cite{akylas2006}.   However the data points reported in \cite{akylas2006} have
been computed  without taking into account possible differences in the
cosmological  evolution properties of the absorbed and unabsorbed AGN
population. This effect, that we have already shown as of the second order for
the HBSS AGN sample, could be  very important for the  \cite{akylas2006} sample
since their objects are at a significantly higher z compared to the HBSS AGN
sample. To test this effect  in Figure \ref{ratio} (middle panel) the
\cite{akylas2006} points  have been rescaled to z=0  using the mean redshift in
each luminosity bins  (T. Akylas, private communication) and a cosmological
evolution of F according to the model  proposed by   \cite{treister2006} ($F
\propto (1+z)^{0.4}$).  A much better agreement with our results is clearly
evident, suggesting a possible  evolution of the fraction of absorbed
AGN with the redshift.   

Finally in Figure \ref{ratio} (right panel) we compare F with that required to
produce the cosmic X-ray background according to the new synthesis model
reported in  \cite{gilli2007} (long dashed line: best fit; short dashed lines:
one $\sigma$ error range; both computed and predicted F ratios refer only to
absorbed AGN with $N_H$ below $10^{24}$ cm$^{-2}$); we discuss this comparison
in section 8.

\section{The intrinsic NH distribution in AGN}
\label{par5}

\begin{figure}
   \centering
\includegraphics[width=6cm]{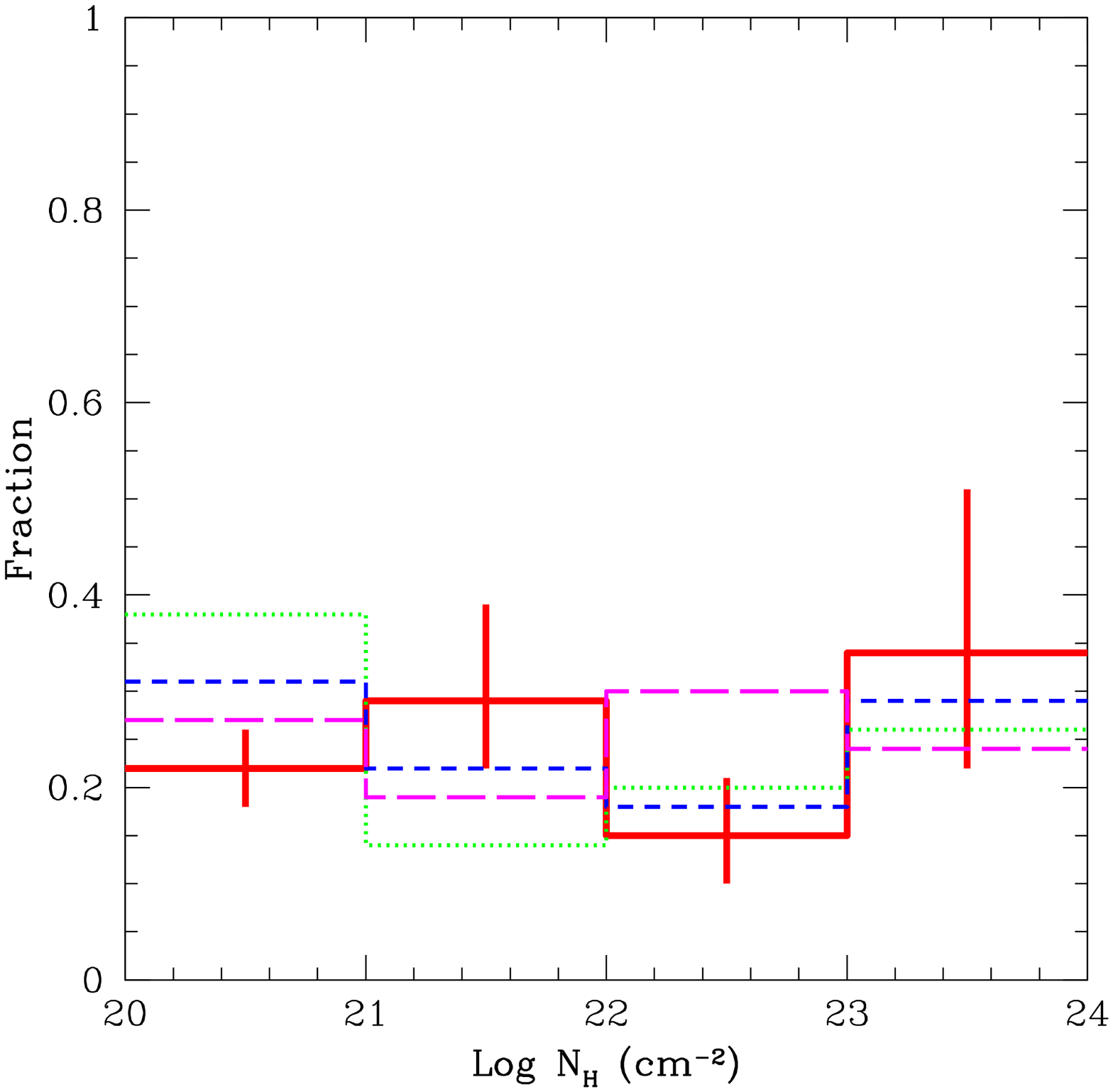}    
   \centering
\includegraphics[width=6cm]{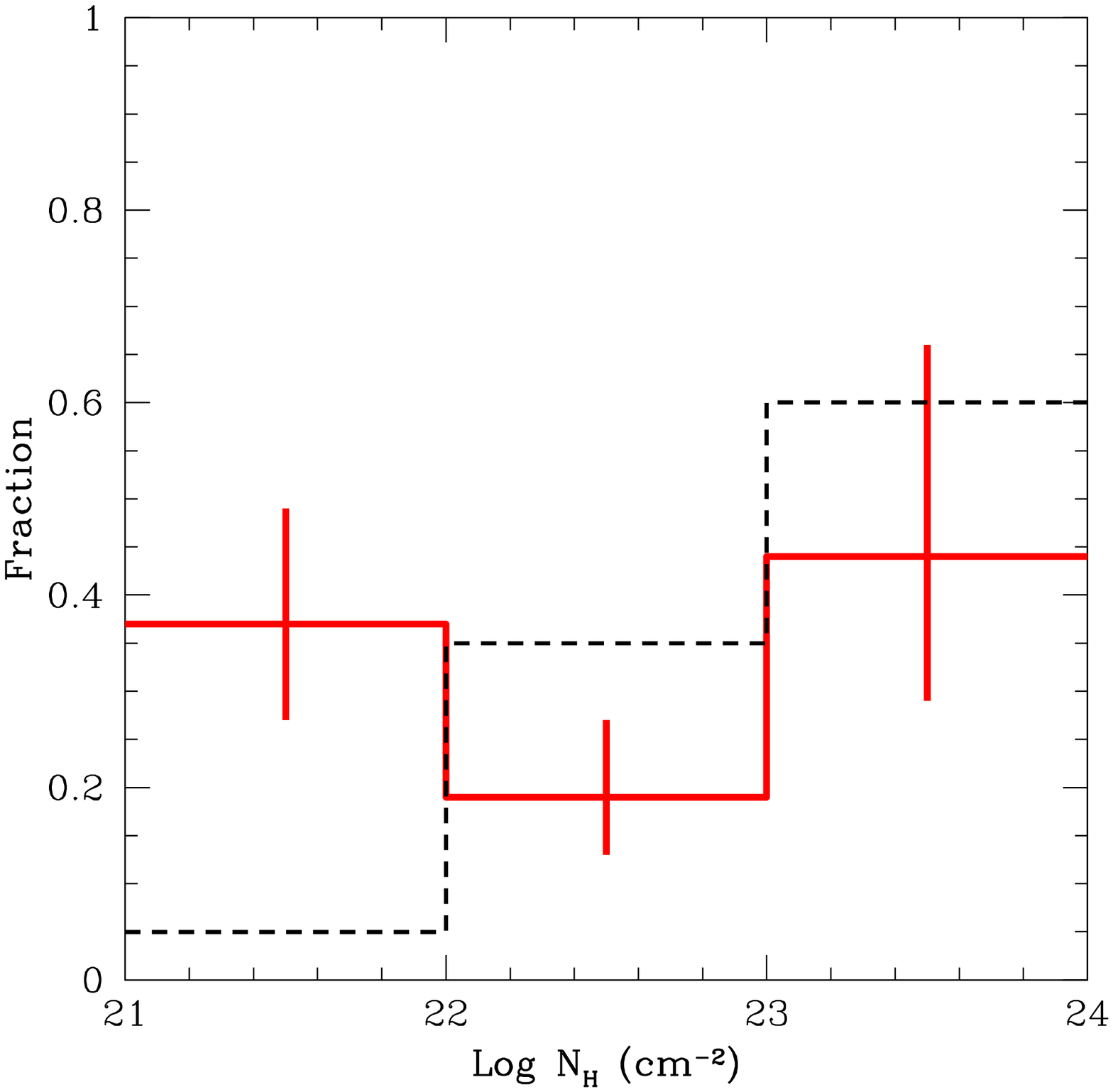}    
  \caption{{\it Upper panel:} intrinsic $N_H$ distribution between $10^{20}$ and
   $10^{24}$ cm$^{-2}$
obtained using the HBSS AGN sample (see section 5 for  details), compared with 
$N_H$ distribution from the SWIFT/BAT sample (\citealt{markwardt2005}: long
dashed line)  and from  the INTEGRAL samples (\citealt{beckmann2006}:  short
dashed line; \citealt{bassani2006}: dotted line). 
{\it Lower panel:} intrinsic $N_H$ distribution between $10^{21}$ and $10^{24}$ cm$^{-2}$
obtained using the HBSS AGN sample, compared with that required from the
synthesis model of the CXB (\cite{gilli2007}). In this panel, as done in 
\citealt{gilli2007}, the number of objects in each $N_H$ bin is normalized to 
the total number of  AGN with $N_H$ between $10^{21}$ and $10^{24}$ cm$^{-2}$.}
         \label{int_nh}
   \end{figure}

A second important and debated issue on the  AGN astrophysics is related to
their intrinsic $N_H$ distribution, i.e. the $N_H$ distribution of the AGN
family computed  taking into account the selection effects related to the
photoelectric  absorption. To compute this distribution we have proceeded as
follows.  We have first split the HBSS AGN sample in four bins of absorbing
column densities up to $N_H$ = $10^{24}$ cm$^{-2}$ 
($N_H$ = $10^{20} - 10^{21}$; $10^{21} - 10^{22}$; $10^{22} - 10^{23}$; $10^{23}
- 10^{24}$ cm$^{-2}$).
Second, we have computed the integral luminosity 
function using the objects in each bin of $N_H$ and the $1/V_a$ method as
discussed in Section 3.2.  Third, using these XLFs we have determined the density
of objects in each $N_H$ bin having a luminosity above $\sim 3\times 10^{42}$
\es (i.e. around the luminosity of the faintest absorbed AGN in the HBSS 
sample).
Finally in order to obtain the $N_H$ distribution (reported in Figure
\ref{int_nh}  upper panel, as solid line)   the density in each  $N_H$ bin has been
normalized to the total density of  AGN with $N_H$ between $10^{20}$ and
$10^{24}$  cm$^{-2}$.  One $\sigma$ Poisson error bars  have been computed using
the \cite{gehrels1986} prescription and the objects in each bin of $N_H$.
It is now interesting to compare the rather flat Log$N_H$ distribution obtained
from the HBSS sample with other samples of AGN or with that required from the
synthesis model of the CXB.
In Figure \ref{int_nh} (upper panel) we compare the $N_H$ distribution derived using
the HBSS sample with that computed using hard ($E>10$ keV) selected (local)
sample of AGN from SWIFT/BAT and INTEGRAL surveys at a flux limit of $\sim 10^{-11}$
\cgs. We stress again that these latter samples are virtually free from
selection bias related to the absorption for $N_H$ below $\simeq 10^{24}$
cm$^{-2}$. Clearly the flat $N_H$ distribution derived using the  HBSS
AGN sample is in agreement with that derived using local samples  of 
X-ray selected AGNat  a flux limit about a factor 100 higher, 
thus suggesting that we are
indeed sampling a similar population of Compton Thin AGN.
The derived flat Log $N_H$ distribution is also consistent with the flat 
Log $N_H$ distribution assumed by \cite{lafranca2005} in the cosmological 
analysis of the HELLAS2XMM AGN sample. 
Finally in Figure \ref{int_nh} (lower panel) we compare the $N_H$ distribution 
derived here with that required by \cite{gilli2007} (up to $N_H=10^{24}$
$cm^{-2}$) to fit the spectra of the CXB; 
the current CXB synthesis modeling based on absorbed and  unabsorbed AGN require
an increasing fraction of absorbed AGN that is clearly at odd with the results
reported here.

It is worth noting that the flat Log $N_H$ distribution derived here is also
markely different from that derived by \cite{risaliti1999} using a sample of
optically selected Seyfert 2 galaxies, with the optically selected sample requiring
an increasing fraction of absorbed AGN with increasing the $N_H$. This difference is
expected and is due to the fact that  we are considering in  Figure 7  both
optically type 1 and optically type 2 AGN, thus providing the $N_H$  distribution 
of the total AGN population. A good agreement with \cite{risaliti1999} is indeed 
obtained is we consider only the HBSS optically Type 2 AGN.  In other words the Log
$N_H$ distribution reported in  \cite{risaliti1999} does not represent the $N_H$
distribution of the total  AGN population but is probably more typical of that of
optically selected Seyfert 2 galaxies, which, with few exceptions, are  absorbed
sources with  $N_H$ usually above $10^{22}$ cm$^{-2}$.

%

\section{Comparison with AGN Unification Models}
\label{par6}

\begin{figure}
   \centering
\includegraphics[width=6cm]{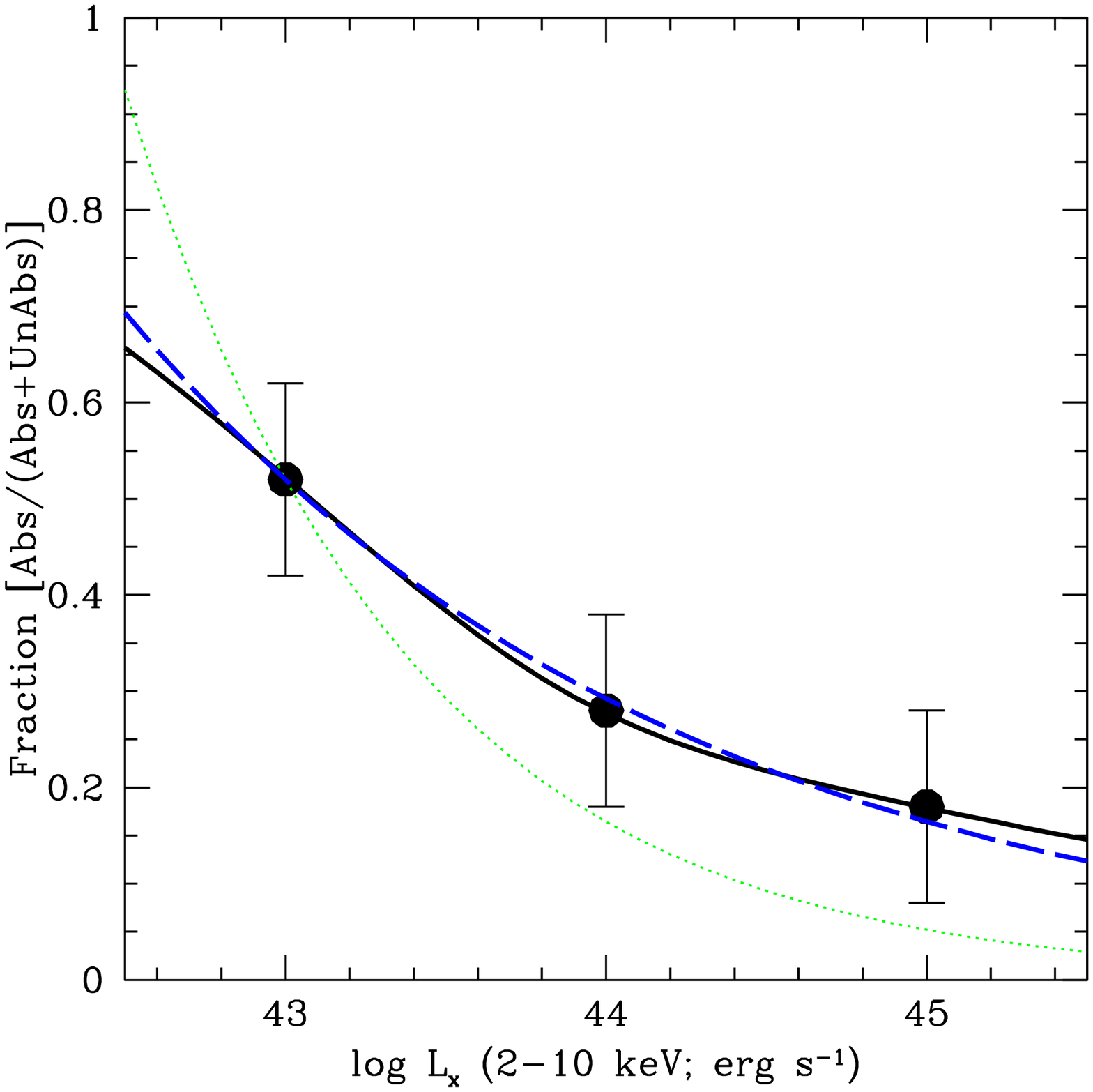}    
   \centering
\includegraphics[width=6cm]{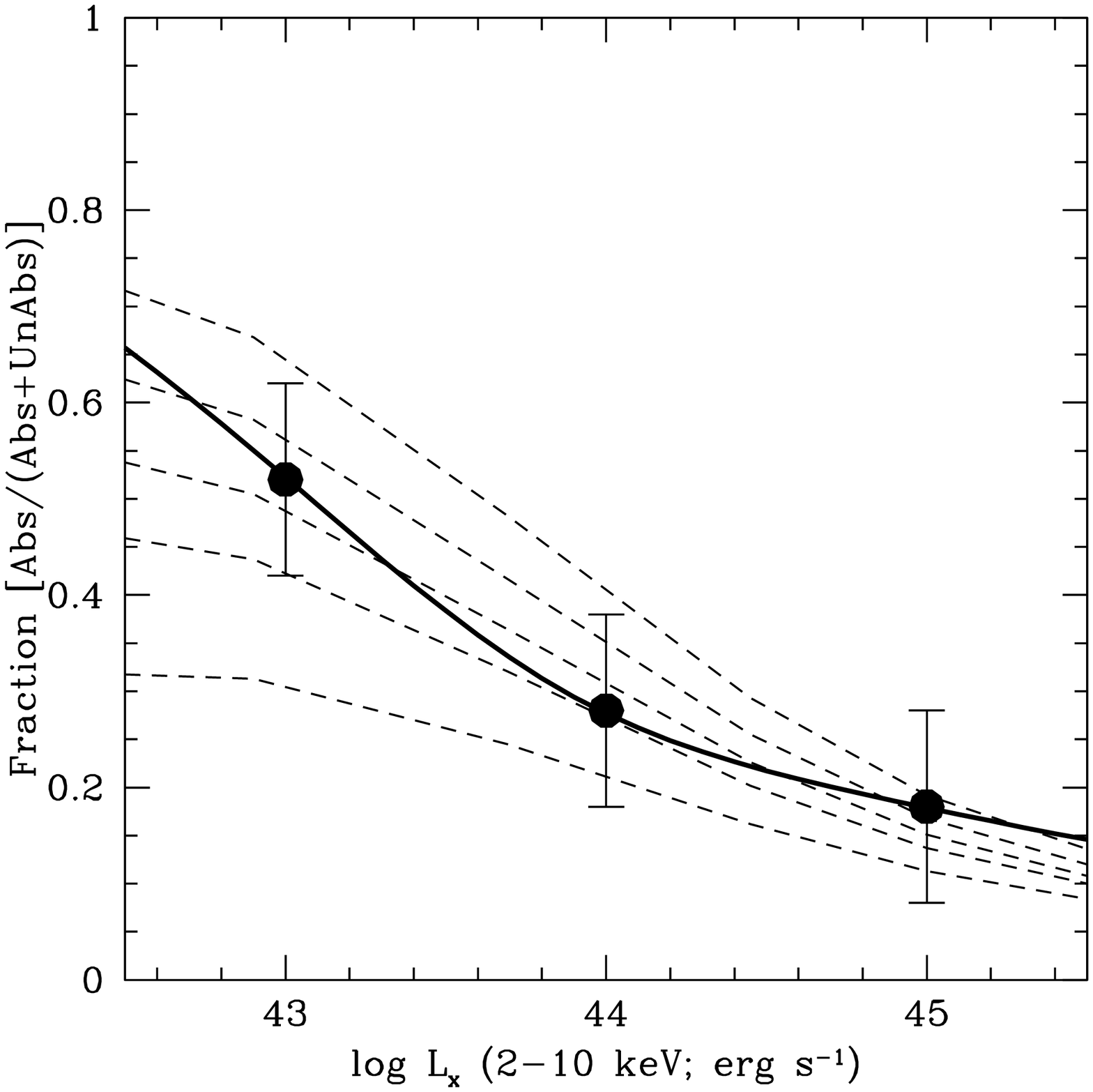}      
  \caption{
{\it Upper panel:} the thick solid line represents the fraction F of absorbed
AGN  as a  function of the intrinsic luminosity obtained using the 
best fit de-evolved
XLF  of absorbed ($N_H$ between $4\times 10^{21}$ and $10^{24}$ cm$^{-2}$)
and unabsorbed ($N_H$ below $4\times 10^{21}$ cm$^{-2}$)
AGN in the HBSS sample.
The 68\%  confidence errors on this ratio are also reported at   $L_x = 10^{43},
10^{44}, 10^{45}$ \es.  The dotted line represents the
absorbed fraction expected from the original  receding torus model discussed in
\cite{lawrence1991} ($F\propto L^{-0.5}$),  while the dashed line represents
the absorbed fraction expected from the  radiation-limited dusty torus recently
discussed in \cite{honig2007} ($F\propto L^{-0.25}$); these expected fractions
have been normalized to the  observed values at $L_x = 10^{43}$ \es;
{\it Lower panel:} the thick solid line as in the upper panel.
The dashed lines represent the
expected behavior of the fraction of absorbed AGN for values of the galaxy
central surface densities (from bottom to top) of 40, 80, 120, 200 and 400
$M_{\odot}/pc^2$ according to the models discussed in \cite{lamastra2006}.
}
         \label{ratio_UM}
   \end{figure}

The results on the fraction of absorbed AGN as a function of the intrinsic
luminosity discussed in section 4 can be now directly compared with the 
prediction(s) of the unification model of AGN. These results are  inconsistent
with the simplest unified scheme, where a similar thick molecular torus (at a
distance of the order of one parsec from the  central black hole) surrounds all
the AGN, and which predicts that F is independent  from L$_x$; a modifications
to this simple zero order unification scheme is clearly required. In Figure
\ref{ratio_UM} (upper panel) we compare F-L$_x$ relationship derived here with
that expected in the framework of the original 
{\it receding torus model}  (F$\propto
L^{-0.5}$, dotted line; \citealt{lawrence1991}) and in the framework of the {\it
radiation-limited clumpy dust torus} (F$\propto L^{-0.25}$, dashed line;
\citealt{honig2007}); both these models  have been normalized with our results 
at $L_x = 10^{43}$ \es.  It is clearly evident the excellent agreement between
our results and the prediction of the radiation-limited dusty torus discussed in
\cite{honig2007}.

Finally in Figure \ref{ratio_UM} (lower panel) we compare the F-L$_x$
relationship  from the HBSS sample with the AGN absorption model recently
discussed by \cite{lamastra2006}.  According to this model the absorption seen
in Compton Thin AGN is not related to the obscuring torus but to molecular gas
in a disk located far away from the X-ray source (between 25 and 450 pc
depending  on the BH mass).  As shown in \cite{lamastra2006} the
anti-correlation between the ratio of absorbed AGN and the luminosity can be
reproduced if a) there is a sufficient number of molecular clouds within the
radius where the BH gravitational influence is dominant and b) there is a
statistical correlation between the BH mass and the  AGN luminosity.
We have reported in Figure \ref{ratio_UM} (lower panel) as  dashed lines  the
expected behavior of the fraction of absorbed AGN for values of the central
surface densities of molecular clouds (from bottom to top) of 40, 80, 120, 200
and 400 $M_{\odot}/pc^2$ as reported in Figure 4 (right panel) of 
\cite{lamastra2006} 
\footnote{A. Lamastra has kindly provided us the expected
behavior of their model for $L_x\gtrsim 3\times 10^{44}$ \es.}, 
based on the
assumption $L_{Bol}/L_{Edd} = 0.1$ and  luminosity dependent-bolometric
corrections (see  \citealt{lamastra2006} for details).
Our results on the F-L$_x$ relationship are fully consistent with this model if
the surface  density of molecular clouds in the central part of the galaxy is
between 80 and 200 $M_{\odot}/pc^2$, a condition that, as discussed in 
\cite{lamastra2006} is consistent with the observations of local galaxies.

%

\section{The X-ray luminosity function  of Compton Thick AGN}
\label{par7}

As discussed above (and as done so far in this paper) XMM-Newton and {\it
Chandra} data can be efficiently  used to investigate the statistical
properties  of  AGN with column densities below  $\sim 10^{24}$ cm$^{-2}$.  On
the contrary  the statistical properties (e.g. XLF) of  absorbed sources  having
$N_H$ above $\sim 10^{24}$ cm$^{-2}$, the so called Compton Thick AGN,
are currently completely unconstrained. According to the latest versions of
synthesis modeling of the  XRB (\citealt{gilli2007}), these sources should
represent a substantial fraction of the total AGN population but only a few
dozens of them (mostly local) have been found and studied so far in the X-ray
domain  (see \citealt{dellaceca2007} for a review). 

\begin{figure}
   \centering
\includegraphics[width=6cm]{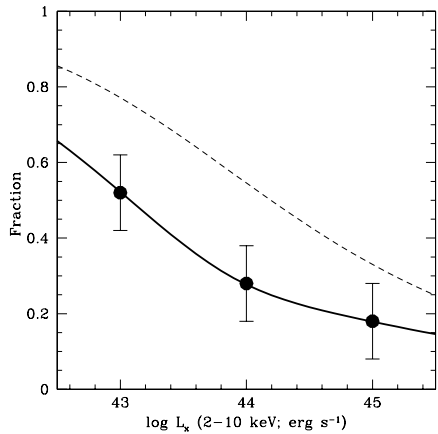}         
  \caption{
The dashed line represents the fraction of optically narrow line AGN 
($F_{optical}$)
as  derived by  \cite{simpson2005} using  a complete, magnitude-limited, sample
of active galaxies from the Sloan  Digital Sky Survey and assuming the mean
$L_{[OIII]}/L_{(2-10 kev)}$ ratio for Seyfert galaxies derived by 
\cite{mulchaey1994}. The thick solid line represents the fraction F of absorbed
AGN (sources with  $N_H$ between $4\times 10^{21}$ and $10^{24}$ cm$^{-2}$
divided by  the sources  with $N_H$ below $10^{24}$ cm$^{-2}$) as a  function of
the intrinsic luminosity obtained using the  best fit de-evolved XLF  of
absorbed  and unabsorbed AGN in the HBSS sample. The 68\%  confidence errors on
this ratio are also reported at  $L_x = 10^{43}, 10^{44}, 10^{45}$ \es.}
       \label{ratio6}
   \end{figure}

In Figure \ref{ratio6} we show  the fraction of optically narrow line AGN as a
function of the X-ray luminosity (dashed line) as derived by \cite{simpson2005}
using  a complete, magnitude-limited, sample of active galaxies from the Sloan 
Digital Sky Survey and assuming the mean  $L_{[OIII]}/L_{(2-10 kev)}\simeq
0.015$ ratio for Seyfert galaxies obtained by \cite{mulchaey1994} (fully
consistent with the similar value reported in \citealt{heckman2005} for the
unobscured view of Seyfert galaxies, $L_{[OIII]}/L_{(2-10 kev)}\simeq 0.017$). 
The thick solid line (and the error bars) are the results derived here on the
fraction of absorbed  ($4\times 10^{21}<$ $N_H$ $\ls 10^{24}$  cm$^{-2}$) AGN
obtained  using the HBSS AGN sample (see Section 4); we recall here  that in the
HBSS  we are sampling only AGN with  $N_H$ below $10^{24}$  cm$^{-2}$.

The behavior of the fraction of optically narrow line AGN  derived by 
\cite{simpson2005} is  remarkable similar to the correlation between F and $L_X$
found here. More important, since the \cite{simpson2005} sample should contain
both Compton Thin ($N_H$ $\ls 10^{24}$  cm$^{-2}$) and Compton Thick  ($N_H$
$\gs 10^{24}$  cm$^{-2}$) AGN (see also the results discussed in 
\citealt{heckman2005}), the comparison between the results found by
\cite{simpson2005} (dashed line in Figure \ref{ratio6};  F$_{optical}$
hereafter) and those found here (solid line in Figure \ref{ratio6}) allow us  to
estimate the ratio  between Compton Thick AGN and  AGN with  $N_H$ between 
$4\times 10^{21}$ and  $\sim 10^{24}$ cm$^{-2}$ as a function of the
luminosity.

%

Assuming that the number of Compton Thick AGN is proportional to the number
of AGN with  $N_H$ in the range  [$4\times 10^{21}; \sim 10^{24}$ cm$^{-2}$], 
$N_{Thick} = C \times N_{4E21-E24}$, thus from the comparision between F and 
F$_{optical}$, and recalling that the definition of F given in Section 4 does
not  include Compton thick AGN in either the numerator or denominator, we can
derive 
$$C = {(F_{optical} -F)\over F\times (1 - F_{optical}).}$$ 

\begin{figure}
   \centering
\includegraphics[width=9cm]{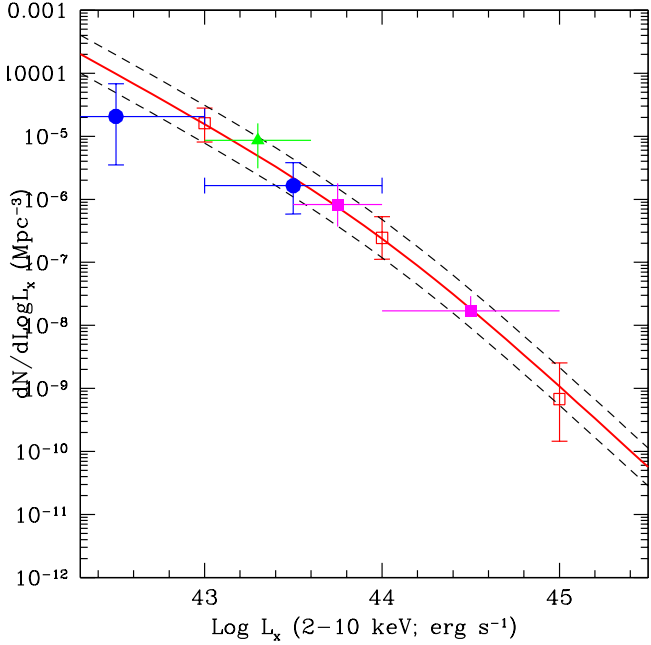}    
  \caption{
The open squares represent the density of Compton Thick AGN at  $L_X = 
10^{43}, 10^{44}$ and  $10^{45}$ \es derived as explained in section 7. The
thick solid line represents the proposed XLF of Compton Thick AGN with 
$A_{Thick} = 2\times A_{4E21-E24}$ ($\gamma_1=1.55$, $\gamma_2=2.61$, Log
$L_{\star} = 44$,  Log A$_{Thick}$ = -6.32) while  the two dashed lines
corresponds to the XLFs with $A_{Thick} = A_{4E21-E24}$ (lower dashed line) and 
$A_{Thick} = 4\times A_{4E21-E24}$ (upper dashed line). The filled circles at
$L_{(2-10 keV)} \sim 3\times 10^{42}$ and  $\sim 3\times 10^{43}$ \es
represent  the density of local Compton Thick AGN derived using the Compton
Thick AGN  in the INTEGRAL all sky survey at $|b|> 5^{\circ}$.  The filled
triangle represents the density of Compton Thick AGN with  $L_X \sim
(1-4)\times 10^{43}$ \es as derived by \cite{daddi2007}.
The filled squares represent the density of Compton Thick AGN  with  $L_X 
\sim (3-10)\times 10^{43}$ \es  and $L_X \sim (1-10)\times 10^{45}$ \es  as
derived by  \cite{fiore2008}.   The densities of Compton Thick AGN reported
in \cite{daddi2007} and  \cite{fiore2008} have been rescaled to z=0 using
our best fit evolutionary model. See section 7 for details.}
         \label{XLF_3}
   \end{figure}

In order to derive an estimate of the density of Compton Thick AGN as a function 
of the luminosity we have first  computed C for three values of the intrinsic
X-ray luminosity ($L_x = 10^{43}$, $L_x = 10^{44}$ and $L_x = 10^{45}$ \es); 
thus we have  multiplied them for the corresponding XLF of  absorbed AGN with 
$N_H$ between  $4\times 10^{21}$ and  $\sim 10^{24}$ cm$^{-2}$ (i.e. second line
of Table 2). The obtained results are reported in Figure \ref{XLF_3} as open
squares (the errors  have been computed taking into account the errors reported
in  Figure \ref{ratio6}). 

These estimates of the density of Compton Thick AGN at  $L_x = 10^{43}$, $L_x =
10^{44}$ and $L_x = 10^{45}$ \es can be well reproduced assuming an XLF similar,
in shape, to that of absorbed AGN reported in Table 2 but having a normalization
$A_{Thick} = 2\times A_{4E21-E24}$;  this XLF is shown as a thick solid line in
Figure \ref{XLF_3} while the two dashed lines corresponds to the XLFs with
$A_{Thick} = A_{4E21-E24}$ (lower dashed line) and   $A_{Thick} = 4\times
A_{4E21-E24}$ (upper dashed line).

%
%

In Figure \ref{XLF_3} we have also reported: 

a) the density of local Compton Thick AGN (filled circles at $L_X \sim 3\times
10^{42}$ and $\sim 3\times 10^{43}$ \es) as derived using the three Compton
Thick AGN (NGC 4945, NGC 3281 and MKN 3) in the INTEGRAL all sky survey at 
$|b|> 5^{\circ}$ (\citealt{sazonov2007})  \footnote{We thank S. Sazonov to have
provided us the sky coverage of this  survey in a tabular form.}. To convert the
observed 17-60 keV fluxes into 2-10 keV intrinsic  (i.e. unabsorbed) 
luminosities we have corrected the fluxes for intrinsic absorption and assumed 
$\Gamma = 1.9$;

b) the density of Compton Thick AGN with  $L_X \sim (1-4)\times 10^{43}$ \es
(filled triangle)  as derived by \cite{daddi2007} using {\it Spitzer} and {\it
Chandra} data in two GOODS fields with  multiwavelenghts information. Their
estimated space density at z between 1.4 and 2.5 has been rescaled to z=0 using
our best fit evolutionary model;

c) the density of Compton Thick AGN (filled squares)  with  $L_X \sim
(3-10)\times 10^{43}$ \es (z between 0.7 and 1.0)  and $L_X \sim (1-10)\times
10^{45}$ \es (z between 1.3 and 2.0)  as derived by  \cite{fiore2008} using  the
{\it Spitzer} and {\it Chandra} data in the COSMOS field. Their estimated space
density  has been rescaled to z=0 using our best fit evolutionary model.

As it is clear in Figure \ref{XLF_3} the different estimates of the space
density of Compton Thick AGN, derived using different seletion criteria (and
methods), are in remarkably good  agreement with the  suggested XLF of Compton
Thick AGN  ($\gamma_1=1.55$, $\gamma_2=2.61$, Log $L_{\star} = 44$,  Log
A$_{Thick}$ = -6.32) with a possible hint of a XLF flattening below few times
$10^{42}$ \es (where, we point out, the XLF of absorbed AGN in the HBSS is not
well  defined since we have no objects in the sample).  

The good agreement seem to suggest that, at the sampled X-ray luminosities, the
space density of optically {\it elusive} Compton Thick AGN  (e.g. objects like
Arp 299, see \citealt{dellaceca2002}) is, at most, similar to  that of optically
selected Compton Thick AGN since the former population should be clearly missing
in the optically selected sample of active galaxies defined in
\cite{simpson2005} but should be present in the infrared or X-ray selected
samples.  This fact is also supported by the consideration that the total number
of Compton Thick AGN cannot be increased arbitrarely but we have to take into
account  the limits imposed by the local black mass density derived by
\cite{marconi2004} from dynamical studies of local galaxy bulges ($\rho_{BH} =
4.0^{1.6}_{1.2} h^2_{65} \times 10^{5}$ M$_{\sun}$ Mpc$^{-3}$ for H$_0$=65). 
Using the formalism described in \cite{lafranca2005} (see their equations  15,
16 and 17), assuming a radiative efficiency $\sim 0.1$, a bolometric  conversion
factor equal to 25 (\citealt{pozzi2007}) and integrating  the XLFs of AGN
(unabsorbed, absorbed and Compton Thick)  from z=4.5 to  z=0 and from  $L_X$ 
$\sim 3\times 10^{42}$ \es to  $\sim 10^{49}$ \es, the limit of  $\rho_{BH} =
5.6 h^2_{65} \times 10^{5}$ M$_{\sun}$ Mpc$^{-3}$ is violated if the XLF of
Compton Thick AGN  is more the $\sim$ 4 times that of absorbed AGN, e.g. the
upper envelope  reported in Figure 10; we stress tha  this is probably an upper
envelope since we have not  considered in the  computation of the local black
mass density the objects with luminosity below  $\sim 3\times 10^{42}$ \es.

Finally the comparison of F$_{optical}$ and F reported in  
Figure \ref{ratio6} can be used to evaluated the ratio, Q, between 
Compton Thick and Compton Thin ($N_H < 10^{24}$ cm$^{-2}$) AGN.
Assuming, as above, that the number of Compton Thick AGN is proportional to 
the number
of AGN with  $N_H$ in the range  [$4\times 10^{21}; \sim 10^{24}$ cm$^{-2}$], thus 
$$Q = {(F_{optical} -F)\over (1 - F_{optical}).}$$

We derive that the density ratio between  Compton Thick AGN and Compton 
Thin AGN decreases from  
$Q=1.08\pm 0.44$ at $\sim 10^{43}$ \es  to 
$Q=0.57\pm 0.22$ at $\sim 10^{44}$ \es  to 
$Q=0.23\pm 0.15$ at $\sim 10^{45}$ \es.  

\section{Summary and Conclusions}
\label{par8}

We have discussed here the cosmological properties of the 62  AGN  belonging to
the  XMM-Hard Bright Serendipitous Survey, a complete and representative  sample
of bright  ($f_X \gtrsim 7\times 10^{-14}$ erg cm$^{-2}$ s$^{-1}$) serendipitous
XMM-Newton sources selected in the $4.5-7.5$ keV energy band on a  sky area of
$\sim 25$ deg$^{-2}$.  Since the HBSS sample is almost completely 
spectroscopically identified  (ID rate $\sim$ 97\%) it allows us to have an
unprecedented and  unbiased view  of the extragalactic  $4.5-7.5$ keV sky in the
bright flux regime.  Using an $N_H$ dividing value of $4\times 10^{21}$
cm$^{-2}$ (see Section 2.1), the HBSS AGN  sample is composed of 40 unabsorbed 
(or marginally absorbed) AGN and 22 absorbed AGN.  

The main results reported and discussed in this paper are:

\begin{enumerate}

\item the HBSS survey is extremely efficient in finding type 2 QSO, i.e. absorbed
AGN with an intrinsic luminosity in excess to  $10^{44}$ \es.  They represent 
$(15\pm 6)\%$ of the total AGN population and $(41\pm 13)\%$ of the  absorbed
ones.   At the flux limit of the HBSS survey  ($\sim 7\times 10^{-14}$ \cgs in
the 4.5--7.5 keV energy range) the measured surface density of type 2 QSOs is  
$0.36^{+0.13}_{-0.12}$ deg$^{-2}$ ($0.28^{+0.15}_{-0.10}$ deg$^{-2}$ if we
consider only the 7 Type 2 QSO with  $N_H > 10^{22}$ cm$^{-2}$).  At the same
flux limit the measured surface density of absorbed (unabsorbed) AGN  is
$0.87^{+0.23}_{-0.18}$  ($1.59^{+0.25}_{-0.25}$) deg$^{-2}$ and the observed 
fraction of absorbed AGN is  ($37\pm 7$)\%;

\item the de-evolved (z=0) 2-10 keV X-ray luminosity function of the total AGN 
population derived using the AGN HBSS sample is in very good agreement with  the
local AGN XLF obtained by \cite{sazonov2004} using the  data from the  Rossi
X-ray  Timing Explorer survey.  A smaller normalization  (a factor between 2.5
to 5 times depending on $L_x$) is observed if compared with the local HEAO1 AGN
XLF as recently computed from  \cite{shinozaki2006};

\item absorbed and unabsorbed AGN are characterized by two different XLF with
the absorbed AGN population being described by a steeper XLF, if compared with
the unabsorbed ones, at all luminosities;

\item the intrinsic fraction of absorbed AGN with $L_X \gtrsim 3\times 10^{42}$
\es (i.e. around the luminosity of the faintest absorbed AGN in the HBSS sample)
is $0.57\pm 0.11$. This is in excellent agreement  with that found using hard
($E>10$ keV) selected (local) sample of AGN from INTEGRAL/Swift surveys  at a
flux limit of $10^{-11}$ \cgs;

\item the fraction F of absorbed AGN is a function of luminosity. A comparison
of our results with those reported in \cite{akylas2006} (obtained  using a
source sample at fainter fluxes and thus at higher z) seem to suggests an 
evolution of the fraction of absorbed AGN with z.  This evolution is consistent
with that recently proposed by \cite{treister2006};

\item we find a flat Log $N_H$ distribution for $N_H$  between $10^{20}$ and 
$10^{24}$  cm$^{-2}$, still in very good agreement with local samples of hard 
($E >$ 10 keV)  selected AGN at a  flux limit of $10^{-11}$ \cgs.  

\item the shape of the F-L$_x$ relationship and the derived flat  Log $N_H$
distribution (both in good agreement with the assumptions and the results of
\citealt{lafranca2005}) are not consistent with the current models for the 
production of the cosmic X-ray background based on the absorbed AGN.  This
disagreement,  probably due to the fact that we are not sampling  the X-ray
source population producing the bulk of the CXB,  may  require a revision in the
current model's parameters, e.g. an increase of the fraction of absorbed sources
with redshift  and/or a $N_H$ distribution which depends on the source 
luminosity, redshift or both;

\item by comparing the results obtained here with the  behavior of the
fraction of optical narrow line AGN  derived by  \cite{simpson2005}  using  a
complete, magnitude-limited, sample of active galaxies from the Sloan  Digital
Sky Survey we have derived, in an indirect way,  the X-ray luminosity function
of Compton Thick AGN; it is described  by a smoothly connected two power-law
function (see section 3.2)  having  $\gamma_1=1.55$, $\gamma_2=2.61$, Log
$L_{\star} = 44$,   Log A$_{Thick}$ = -6.32.  The density ratio, Q,  between 
Compton Thick AGN ($N_H \gs 10^{24}$  cm$^{-2}$) and Compton Thin AGN ($N_H \ls
10^{24}$  cm$^{-2}$) decreases from   $Q=1.08\pm 0.44$ at $\sim 10^{43}$ \es 
to  $Q=0.57\pm 0.22$ at $\sim 10^{44}$ \es  to  $Q=0.23\pm 0.15$ at $\sim
10^{45}$ \es;

\item finally, the decreasing fraction of absorbed AGN as a function of the
luminosity is fully consistent with the hypothesis of a reduction of the
covering factor of the {\it gas} component as a function of the luminosity.  A
similar reduction with the luminosity of the covering factor of the circumnuclear  {\it dust}
component  has been recently pointed  out by \cite{maiolino2007} using a sample
of AGN observed with {\it Spitzer}. These results are clearly inconsistent with
the simplest unified scheme of AGN.   Indeed we found an  excellent agreement
between our results and the predictions of the radiation-limited dusty torus
discussed in \cite{honig2007}. An alternative model which also  explains well
our results has been  recently discussed by \cite{lamastra2006} and is based on
the obscuration  due to molecular gas in a disk located far away from the X-ray
source (between 25 and 450 pc depending on the BH mass).

\end{enumerate}

\begin{acknowledgements}


Based on observations made with: 
ESO Telescopes at the La Silla  and Paranal Observatories; 
the Italian Telescopio Nazionale Galileo (TNG) operated on the island of 
  La Palma by the Fundaci\'on Galileo Galilei of the INAF (Istituto Nazionale di 
  Astrofisica);
the Spanish Observatorio del Roque de los Muchachos of the Instituto de 
  Astrofisica de Canarias; 
the German-Spanish Astronomical Center, Calar Alto (operated jointly by 
  Max-PlanckInstitut  f\"{u}r Astronomie and Instututo de Astrofisica de 
  Andalucia, CSIC);
XMM-Newton, an ESA science mission with instruments and contributions directly
funded by ESA Member States and the USA (NASA).
We thanks A. Lamastra, T.Akylas, R. Gilli and S. Sazonov to have provided 
us their results 
in a tabular form and A. Comastri, R. Gilli, F. La Franca and C. Perola for
useful comments.  Finally we sincerely thanks the anonymous 
referee for the useful
comments that have substantially improved the paper.
RDC, AC, TM and PS acknowledge financial support from the  MIUR, grant PRIN-MUR
2006-02-5203 and from the Italian Space Agency (ASI), grants n. I/088/06/0 and n.
I/023/05/0. 
FJC acknowledges financial support by the Spanish Ministry of Education and
Science, through project ESP2006-13608-C01-01.
This research has made use of the Simbad database and of the  NASA/IPAC
Extragalactic Database (NED) which is operated by the Jet Propulsion Laboratory,
California Institute of Technology, under contract with the National Aeronautics
and Space Administration.

\end{acknowledgements}


\addtocounter{table}{-3}
\onecolumn
\begin{longtable}{l r c l c r c}
\caption{Basic X-ray information of the XMM-{\it Newton} HBSS AGN Sample}\\
\hline\hline
Source           &  Rate                   & z   &
$\Gamma$          & $N_H$                          &   
f$_{2-10 keV}$   &  Log L$_{2-10 keV}$ \\
                 & $\times 10^{-3}$        &      &   
                 & $\times 10^{22}$               &                         
$\times 10^{-13}$ &                   \\
	         &  cts/s                 &      &
                 &  cm$^{-2}$                    &                         
erg cm$^{-2}$ s$^{-1}$    & erg s$^{-1}$ \\
(1) & (2) & (3) & (4) & (5) & (6) & (7)\\  
\hline
\endfirsthead
\caption{continued.}\\
\hline\hline
Source           &  Rate                   & z   &
$\Gamma$          & $N_H$                          &   
f$_{2-10 keV}$   &  Log L$_{2-10 keV}$ \\
                 & $\times 10^{-3}$        &      &   
                 & $\times 10^{22}$               &                         
$\times 10^{-13}$ &                   \\
	         &  cts/s                 &      &
                 &  cm$^{-2}$                    &                         
erg cm$^{-2}$ s$^{-1}$    & erg s$^{-1}$ \\
(1) & (2) & (3) & (4) & (5) & (6) & (7)\\  
\hline
\endhead
\hline
\endfoot
 XBSJ002618.5+105019	 &   2.34  $\pm$  0.55   &  0.473   &    2.02 $_{ -0.06 }^{+ 0.07}$  & $<$  0.011                         &   2.83   &  44.44   \\ 
 XBSJ013240.1$-$133307	  &   3.23  $\pm$  0.71   &  0.562   &    1.90 (frozen)              &      2.540 $_{ -0.560 }^{+ 0.710}$ &   1.76   &  44.43   \\ 
 XBSJ013944.0$-$674909	  &   2.05  $\pm$  0.46   &  0.104   &    1.94 $_{ -0.12 }^{+ 0.14}$  & $<$  0.018                         &   1.15   &  42.56   \\ 
 XBSJ015957.5+003309	 &   3.80  $\pm$  0.92   &  0.310   &    2.14 $_{ -0.09 }^{+ 0.09}$  & $<$  0.010                         &   2.91   &  44.03   \\ 
 XBSJ021640.7$-$044404	  &   2.08  $\pm$  0.49   &  0.873   &    2.24 $_{ -0.09 }^{+ 0.08}$  & $<$  0.020                         &   1.11   &  44.74   \\ 
 XBSJ021808.3$-$045845	  &   2.67  $\pm$  0.25   &  0.712   &    2.01 $_{ -0.04 }^{+ 0.04}$  & $<$  0.007                         &   2.32   &  44.79   \\ 
 XBSJ021817.4$-$045113	  &   3.30  $\pm$  0.40   &  1.080   &    1.84 $_{ -0.03 }^{+ 0.04}$  & $<$  0.040                         &   2.65   &  45.22   \\ 
 XBSJ021822.2$-$050615	  &   4.54  $\pm$  0.43   &  0.044   &    1.66 $_{ -0.36 }^{+ 0.34}$  &     20.540 $_{ -0.440 }^{+ 0.360}$ &   3.00   &  42.53   \\ 
 XBSJ023713.5$-$522734	  &   3.23  $\pm$  0.61   &  0.193   &    1.90 $_{ -0.13 }^{+ 0.18}$  & $<$  0.180                         &   2.74   &  43.52   \\ 
 XBSJ030206.8$-$000121	  &   3.10  $\pm$  0.42   &  0.641   &    1.90 $_{ -0.06 }^{+ 0.06}$  & $<$  0.015                         &   2.23   &  44.63   \\ 
 XBSJ030614.1$-$284019	  &   4.61  $\pm$  0.92   &  0.278   &    1.56 $_{ -0.13 }^{+ 0.26}$  & $<$  0.048                         &   2.91   &  43.87   \\ 
 XBSJ031015.5$-$765131	  &   4.39  $\pm$  0.46   &  1.187   &    1.92 $_{ -0.04 }^{+ 0.04}$  & $<$  0.022                         &   3.41   &  45.47   \\ 
 XBSJ031146.1$-$550702	  &   5.87  $\pm$  1.12   &  0.162   &    2.08 $_{ -0.12 }^{+ 0.13}$  & $<$  0.013                         &   2.79   &  43.37   \\ 
 XBSJ031859.2$-$441627	  &   2.16  $\pm$  0.49   &  0.140   &    1.72 $_{ -0.37 }^{+ 0.43}$  &      0.390 $_{ -0.280 }^{+ 0.340}$ &   1.63   &  42.99   \\ 
 XBSJ033845.7$-$352253	  &   2.37  $\pm$  0.36   &  0.113   &    1.90 (frozen)              &     28.500 $_{-10.000 }^{+ 6.700}$ &   1.67   &  43.21   \\ 
 XBSJ040658.8$-$712457	  &   3.41  $\pm$  0.68   &  0.181   &    1.90 (frozen)              &     21.970 $_{-12.780 }^{+17.670}$ &   1.59   &  43.56   \\ 
 XBSJ040758.9$-$712833	  &   4.96  $\pm$  0.91   &  0.134   &    1.90 (frozen)              &     28.230 $_{-14.640 }^{+21.960}$ &   2.56   &  43.44   \\ 
 XBSJ041108.1$-$711341	  &   2.20  $\pm$  0.42   &  0.923   &    1.95 $_{ -0.36 }^{+ 0.48}$  & $<$  0.610                         &   0.82   &  44.59   \\ 
 XBSJ050536.6$-$290050	  &   2.19  $\pm$  0.44   &  0.577   &    1.85 $_{ -0.18 }^{+ 0.20}$  &      0.610 $_{ -0.170 }^{+ 0.210}$ &   1.33   &  44.27   \\ 
 XBSJ052108.5$-$251913	  &   2.11  $\pm$  0.55   &  1.196   &    1.72 $_{ -0.19 }^{+ 0.29}$  &      0.100 $_{ -0.100 }^{+ 1.800}$ &   2.46   &  45.26   \\ 
 XBSJ052128.9$-$253032	  &   3.08  $\pm$  0.88   &  0.588   &    1.90 (frozen)              &     12.710 $_{ -3.980 }^{+ 6.460}$ &   1.34   &  44.45   \\ 
 XBSJ074202.7+742625	 &   3.38  $\pm$  0.49   &  0.599   &    2.02 $_{ -0.15 }^{+ 0.16}$  &      0.070 $_{ -0.060 }^{+ 0.070}$ &   1.64   &  44.46   \\ 
 XBSJ074312.1+742937	 &  10.92  $\pm$  0.64   &  0.312   &    1.99 $_{ -0.07 }^{+ 0.06}$  & $<$  0.040                         &   9.67   &  44.55   \\ 
 XBSJ083737.0+255151	 &   2.98  $\pm$  0.71   &  0.105   &    1.77 $_{ -0.33 }^{+ 0.35}$  &      0.340 $_{ -0.220 }^{+ 0.300}$ &   3.04   &  43.00   \\ 
 XBSJ083737.1+254751	 &   7.30  $\pm$  1.12   &  0.080   &    1.92 $_{ -0.12 }^{+ 0.14}$  &      0.150 $_{ -0.040 }^{+ 0.040}$ &   6.63   &  43.09   \\ 
 XBSJ091828.4+513931	 &   3.03  $\pm$  0.54   &  0.185   &    1.47 $_{ -0.30 }^{+ 0.51}$  &      4.240 $_{ -1.410 }^{+ 3.030}$ &   2.62   &  43.52   \\ 
 XBSJ095218.9$-$013643	  &  24.50  $\pm$  2.98   &  0.020   &    1.90 (frozen)              &     29.400 $_{ -7.000 }^{+ 9.600}$ &  11.00   &  42.54   \\ 
 XBSJ101850.5+411506	 &   2.16  $\pm$  0.49   &  0.577   &    2.29 $_{ -0.05 }^{+ 0.09}$  & $<$  0.025                         &   1.51   &  44.43   \\ 
 XBSJ101922.6+412049	 &   2.46  $\pm$  0.47   &  0.239   &    2.13 $_{ -0.10 }^{+ 0.28}$  & $<$  0.040                         &   2.87   &  43.71   \\ 
 XBSJ104026.9+204542	 &   8.70  $\pm$  1.15   &  0.465   &    1.99 $_{ -0.05 }^{+ 0.05}$  & $<$  0.009                         &   6.21   &  44.76   \\ 
 XBSJ104522.1$-$012843	  &   3.00  $\pm$  0.67   &  0.782   &    2.05 $_{ -0.11 }^{+ 0.11}$  & $<$  0.038                         &   2.11   &  44.85   \\ 
 XBSJ104912.8+330459	 &   2.06  $\pm$  0.51   &  0.226   &    1.68 $_{ -0.17 }^{+ 0.19}$  &      0.030 $_{ -0.030 }^{+ 0.050}$ &   1.83   &  43.48   \\ 
 XBSJ112026.7+431520	 &   2.32  $\pm$  0.37   &  0.146   &    1.65 $_{ -0.42 }^{+ 0.83}$  &      6.280 $_{ -1.730 }^{+ 3.210}$ &   1.87   &  43.20   \\ 
 XBSJ113106.9+312518	 &   2.27  $\pm$  0.44   &  1.482   &    1.77 $_{ -0.26 }^{+ 0.37}$  &      0.065 $_{ -0.065 }^{+ 0.620}$ &   1.08   &  45.14   \\ 
 XBSJ113121.8+310252	 &   3.88  $\pm$  0.69   &  0.190   &    1.62 $_{ -0.28 }^{+ 0.10}$  &      0.760 $_{ -0.300 }^{+ 0.250}$ &   2.70   &  43.50   \\ 
 XBSJ113148.7+311358	 &   3.43  $\pm$  0.48   &  0.500   &    1.90 (frozen)              &      3.200 $_{ -0.460 }^{+ 0.540}$ &   2.96   &  44.54   \\ 
 XBSJ122656.5+013126	 &   3.07  $\pm$  0.54   &  0.733   &    1.61 $_{ -0.21 }^{+ 0.23}$  &      2.390 $_{ -0.550 }^{+ 0.650}$ &   2.23   &  44.73   \\ 
 XBSJ124641.8+022412	 &   2.22  $\pm$  0.49   &  0.934   &    2.22 $_{ -0.10 }^{+ 0.10}$  & $<$  0.038                         &   1.36   &  44.90   \\ 
 XBSJ132038.0+341124	 &   2.89  $\pm$  0.40   &  0.065   &    1.77 $_{ -0.10 }^{+ 0.13}$  &      0.150 $_{ -0.030 }^{+ 0.040}$ &   2.54   &  42.48   \\ 
 XBSJ133942.6$-$315004	  &   3.52  $\pm$  0.51   &  0.114   &    1.62 $_{ -0.18 }^{+ 0.22}$  &      0.230 $_{ -0.080 }^{+ 0.120}$ &   1.70   &  42.81   \\ 
 XBSJ134656.7+580315	 &   3.33  $\pm$  0.56   &  0.373   &    1.90 (frozen)               &      9.500 $_{ -3.160 }^{+ 4.720}$ &   1.47   &  44.04   \\ 
 XBSJ134749.9+582111	 &   7.39  $\pm$  0.66   &  0.646   &    2.19 $_{ -0.03 }^{+ 0.02}$  & $<$  0.015                         &   5.14   &  45.07   \\ 
 XBSJ140102.0$-$111224	  &   7.21  $\pm$  0.59   &  0.037   &    1.90 $_{ -0.02 }^{+ 0.04}$  & $<$  0.004                         &   5.21   &  42.28   \\ 
 XBSJ140113.4+024016	 &   2.10  $\pm$  0.52   &  0.631   &    2.05 $_{ -0.24 }^{+ 0.42}$  & $<$  0.170                         &   0.34   &  43.83   \\ 
 XBSJ140127.7+025605	 &   6.66  $\pm$  0.62   &  0.265   &    1.57 $_{ -0.06 }^{+ 0.05}$  &      0.150 $_{ -0.020 }^{+ 0.020}$ &   7.42   &  44.23   \\ 
 XBSJ141531.5+113156	 &   2.58  $\pm$  0.34   &  0.257   &    1.84 $_{ -0.05 }^{+ 0.07}$  & $<$  0.022                         &   2.09   &  43.68   \\ 
 XBSJ142741.8+423335	 &   3.62  $\pm$  0.53   &  0.142   &    1.90 (frozen)              &      4.480 $_{ -0.770 }^{+ 0.930}$ &   2.11   &  43.20   \\ 
 XBSJ143835.1+642928	 &   3.57  $\pm$  0.54   &  0.118   &    1.84 $_{ -0.18 }^{+ 0.42}$  &      1.850 $_{ -0.550 }^{+ 0.730}$ &   2.33   &  43.03   \\ 
 XBSJ143911.2+640526	 &   2.44  $\pm$  0.39   &  0.113   &    1.90 (frozen)              &     20.000 $_{ -6.560 }^{+ 9.720}$ &   1.09   &  42.96   \\ 
 XBSJ153452.3+013104	 &   8.36  $\pm$  1.30   &  1.435   &    1.75 $_{ -0.04 }^{+ 0.09}$  & $<$  0.050                         &   7.56   &  45.95   \\ 
 XBSJ160645.9+081525	 &   7.36  $\pm$  1.37   &  0.618   &    1.75 $_{ -0.71 }^{+ 0.71}$  &     14.270 $_{ -5.610 }^{+ 6.290}$ &   3.84   &  44.94   \\ 
 XBSJ161820.7+124116	 &   2.09  $\pm$  0.61   &  0.361   &    1.90 (frozen)              &      4.960 $_{ -2.480 }^{+ 6.140}$ &   0.88   &  43.72   \\ 
 XBSJ165425.3+142159	 &   5.26  $\pm$  1.08   &  0.178   &    2.11 $_{ -0.03 }^{+ 0.07}$  & $<$  0.016                         &   6.34   &  43.83   \\ 
 XBSJ165448.5+141311	 &   6.25  $\pm$  1.79   &  0.320   &    1.84 $_{ -0.08 }^{+ 0.13}$  & $<$  0.043                         &   4.46   &  44.22   \\ 
 XBSJ193248.8$-$723355	  &   4.47  $\pm$  0.73   &  0.287   &    1.48 $_{ -0.22 }^{+ 0.25}$  &      0.730 $_{ -0.290 }^{+ 0.350}$ &   2.30   &  43.80   \\ 
 XBSJ204043.4$-$004548	  &   3.24  $\pm$  0.80   &  0.615   &    1.90 (frozen)              &      3.000 $_{ -0.900 }^{+ 1.150}$ &   1.66   &  44.50   \\ 
 XBSJ205635.7$-$044717	  &   2.08  $\pm$  0.50   &  0.217   &    1.91 $_{ -0.35 }^{+ 0.51}$  & $<$  0.560                         &   1.66   &  43.42   \\ 
 XBSJ205829.9$-$423634	  &   3.91  $\pm$  0.56   &  0.232   &    1.91 $_{ -0.09 }^{+ 0.09}$  &      0.090 $_{ -0.030 }^{+ 0.040}$ &   3.17   &  43.76   \\ 
 XBSJ213002.3$-$153414	  &   2.30  $\pm$  0.47   &  0.562   &    2.10 $_{ -0.21 }^{+ 0.23}$  &      0.080 $_{ -0.080 }^{+ 0.120}$ &   1.84   &  44.45   \\ 
 XBSJ213820.2$-$142536	  &   2.80  $\pm$  0.59   &  0.369   &    1.61 $_{ -0.13 }^{+ 0.13}$  &      0.510 $_{ -0.120 }^{+ 0.130}$ &   2.26   &  44.05   \\ 
 XBSJ214041.4$-$234720	  &   3.30  $\pm$  0.68   &  0.490   &    2.19 $_{ -0.10 }^{+ 0.10}$  & $<$  0.012                         &   1.69   &  44.28   \\ 
 XBSJ220601.5$-$015346	  &   2.26  $\pm$  0.55   &  0.211   &    1.56 $_{ -0.09 }^{+ 0.16}$  & $<$  0.670                         &   1.66   &  43.36   \\ 
\hline\hline
\end{longtable}

Columns are as follows:
(1) Source name; 
(2) Source MOS2 count rate, and $1\sigma$ error, in the 4.5-7.5 keV energy band
   (units of $10^{-3}$ cts/s). Please note that the reported count rates have 
    been corrected for vignetting;
(3) Redshift;
(4) Source photon index, and $90\%$ confidence errors;
(5) Absorbing column density at the source redshift, 
    and $90\%$ confidence errors;
(6) Observed 2-10 keV flux corrected for Galactic absorption;
(7) Intrinsic (i.e. de-absorbed)  2--10 keV luminosity.
\twocolumn

\appendix

\section{Computation of the intrinsic 2-10 keV XLF using the HBSS AGN sample}

In Section 3 we have derived the intrinsic 2-10 keV XLF of absorbed and 
unabsorbed AGN using the HBSS AGN sample selected in the 4.5-7.5 keV energy
band. To this purpose we have used the $1/V_a$ method, a well established 
procedure to derive space densities of interesting astronomical objects 
(\citealt{avni1980}).

However to apply this method to the absorbed AGN population the following
considerations have to be taken into account. Given an intrinsic count rate
(i.e. the count rate  we would have observed without  intrinsic
absorption), the  observed count rate is a function of the absorbing column
density $N_H$  (at the source redshift) and of the  energy selection band.
Therefore, a given source with an intrinsic count rate  above the survey
threshold could have been missed in the survey since, because of the absorption,
its observed count rate falls below the survey count rate threshold. 

   \begin{figure}
   \centering
\includegraphics[width=7cm]{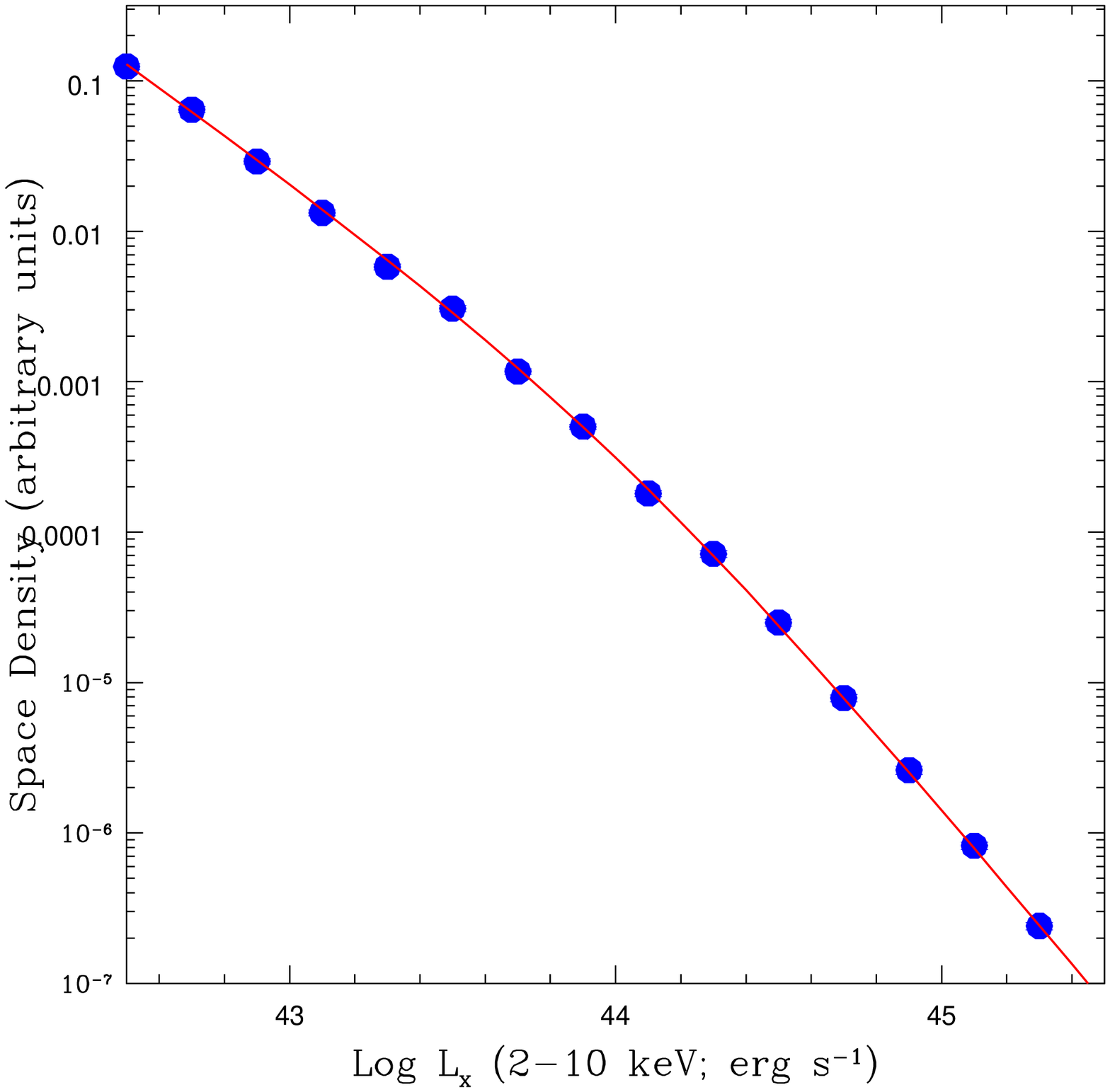}    
   \centering
\includegraphics[width=7cm]{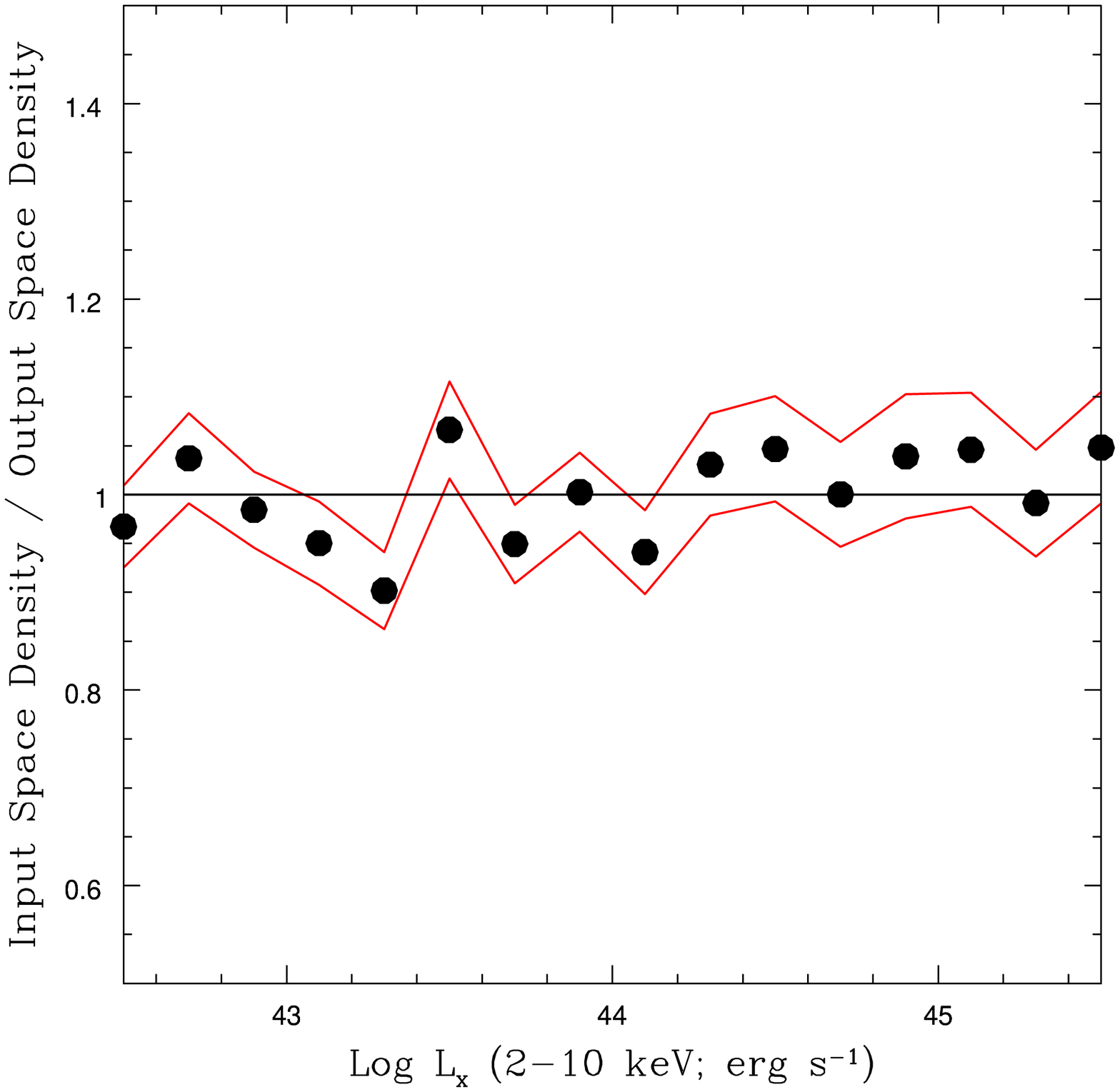}    
  \caption{The results reported in this Figure are those relative to the sources
  filtered by an absorbing column density $N_H$ between $10^{23}$ and 
  $10^{24}$ cm$^{-2}$. {\it Upper panel}: comparison between the input XLF
  (solid line) and the XLF computed (dots) using the method discussed here. 
  {\it Lower panel}: ratio between the input and output XLF. The solid lines
  enclose 1$\sigma$ error bars on this ratio.}
         \label{simul}
   \end{figure}
%

In order to compute the $z_{max}$ of each source (absorbed or not) in section 3
we have used the observed MOS2  count rate in the selection band (4.5-7.5 keV)
 and
the count rate limit of the HBSS survey in the same energy band ($=2\times
10^{-3}$ cts/s). The intrinsic (2-10 keV) luminosities (as derived from the
X-ray spectral  analysis of each source) have been thus used  to divide the
sources into luminosity bins and to derive the luminosity  function.  A similar
procedure was recently used by  \cite{shinozaki2006} in computing the local
luminosity function of the absorbed and unabsorbed AGN in the HEAO1 surveys; 
however \cite{shinozaki2006} did not prove the correctness of this approach and,
in particular,  if this approach may recover the {\it true} space density of
the absorbed AGN population. 

To investigate  this aspect we have simulated a source sample assuming:

a) a 2-10 keV local luminosity function described by a smoothly connected two
power-law  function (see section 3.2) having the following parameters
$\gamma_1=1.55$, $\gamma_2=2.61$, LogL$^*=44$.  These parameters are those
describing the intrinsic   2-10 keV XLF of the absorbed AGN population in the
HBSS survey;

b) a LDDE cosmological evolution model (see section 3.1)  having  best fit
parameters as follows: p1=6.5, p2=-1.15, z$_{\rm c}^*=2.49$, $\alpha$=0.20, Log
L$_a$=45.80;

c) a flat intrinsic Log N$_H$ distribution from $10^{20}$ to $10^{25}$  cm$^{-2}$.

Folding all this information together we have produced a simulated sample of
$\sim$ 40000 sources that would be visible at the count rate threshold of the
HBSS survey.  For each source we have computed  the observed count rate  in the
4.5-7.5 keV, the redshift, the absorbing column density N$_H$ and the intrinsic
luminosity in the 2-10 keV band. For simplicity we have assumed the same
intrinsic photon  index ($\Gamma=1.9$) for all the sources.  This simulated
sample has been thus analyzed with the same procedure used for the real
HBSS source sample, namely, using  the observed 4.5-7.5 count rate  (and the
count rate limit of the HBSS survey), combined with the input LDDE model,  to
evaluate space densities  and the intrinsic (2-10 keV) luminosities to split
the sources into luminosity bins.

The results obtained for one extreme example (the sources having an absorbing
column density in the range between $10^{23}$  and $10^{24}$ cm$^{-2}$) are
shown in Figure A.1;  we have reported the input XLF, the output XLF and the
ratio between the derived XLF at z=0 from the simulated sample and the input
local XLF. Is is worth noting the very good agreement between the input and
output XLF. Similar results have been obtained for less extreme values  of the
$N_H$. For the purpose of this paper the main conclusion we reached using these 
simulations is that for the  HBSS selection band (4.5-7.5 keV) our approach to
compute  space densities can recover the {\it true}  space density of the AGN
population as a function of the intrinsic $N_H$  up to $N_H \sim  10^{24}$
cm$^{-2}$,  {\bf even if}  in the $N_H$ bin between $10^{23}$  and $10^{24}$
about 50 \% of the AGN are  lost (i.e. observed count rate is below the survey
threshold) because of the intrinsic absorption.   Our simulations also show
that  the above statement is not true for $N_H$ above $10^{24}$ where a dramatic
number of objects  ($\sim 95$\% between $10^{24}$  and $10^{25}$) are lost
because of absorption (note that for these high  $N_H$ values the situation
could be even worst if Compton scattering is considered). In summary using the
HBSS survey and our computation method we can recover the true space density  of
absorbed AGN up to $N_H \sim 10^{24}$ cm$^{-2}$, while we are quite  inefficient
in selecting AGN with an intrinsic $N_H$  above this value.  The shape of the
intrinsic XLF, at least for AGN with  $N_H$  up to $\sim 10^{24}$ cm$^{-2}$,  is
therefore recovered allowing  us to compare the XLF of absorbed and unabsorbed
AGN consistently. We have repeated these simulations using different values of
the parameters for the XLF and/or for the cosmological evolution  obtaining the
same results thus giving us confidence about the overall procedure.

\end{document}